\documentclass[11pt]{amsart}
\usepackage[utf8]{inputenc}
\usepackage[titletoc]{appendix}

\usepackage[backend=biber, style=numeric, date=year, url=false, doi=true, isbn=false, eprint=true, giveninits=true, maxcitenames=5, maxbibnames=5]{biblatex}
\AtEveryBibitem{%
  \ifentrytype{online}{%
  }{%
    \clearfield{eprint}%
    \clearfield{urldate}%
  }%
}
\usepackage{blindtext}
\usepackage{amssymb}
\usepackage{amsfonts}
\usepackage{amsthm}
\usepackage{tikz}
\usepackage{caption}
\usepackage{graphics}
\usepackage{xcolor}
\usepackage{shadethm}
\usepackage{subfigure}
\usepackage{epigraph}
\usepackage[includeheadfoot,margin=1.2in]{geometry}
\usepackage{placeins}
\usetikzlibrary{snakes}
\usepackage[new]{old-arrows}

\usepackage{enumerate}
\usepackage{verbatim}
\usetikzlibrary{trees}
\usetikzlibrary{decorations.markings}
\usetikzlibrary{arrows}
\usetikzlibrary{cd}
\usepackage{mathrsfs}
\usepackage{xparse}
\usetikzlibrary{positioning,arrows,patterns}
\usetikzlibrary{decorations.markings}
\usetikzlibrary{calc}
\tikzset{
  photon/.style={decorate, decoration={snake}, draw=black},
  fermion/.style={draw=black, postaction={decorate},decoration={markings,mark=at position .55 with {\arrow{>}}}},
  fermion2/.style={dashed, dash phase=0.1pt, draw=black, postaction={decorate},decoration={markings,mark=at position .55 with {\arrow{>}}}},
  vertex/.style={draw,shape=circle,fill=black,minimum size=5pt,inner sep=0pt},
particle/.style={thick,draw=black},
particle2/.style={thick,draw=blue},
avector/.style={thick,draw=black, postaction={decorate},
    decoration={markings,mark=at position 1 with {\arrow[black]{triangle 45}}}},
gluon/.style={decorate, draw=black,
    decoration={coil,aspect=0}}
 }
\NewDocumentCommand\semiloop{O{black}mmmO{}O{above}}
{%
\draw[#1] let \p1 = ($(#3)-(#2)$) in (#3) arc (#4:({#4+180}):({0.5*veclen(\x1,\y1)})node[midway, #6] {#5};)
}

\usepackage[mathcal]{euscript}

\usepackage{url}
\usepackage{hyperref}
\hypersetup{colorlinks=true, citecolor=purple, linktocpage, linkcolor=blue, urlcolor=cyan} 

\theoremstyle{plain}
\newtheorem{thm}{Theorem}[section]

\theoremstyle{definition}

\newtheorem*{thm*}{Theorem}
\newtheorem*{lem*}{Lemma}
\newtheorem*{prop*}{Proposition}
\newtheorem*{cor*}{Corollary}
\newtheorem*{exe*}{Exercise}
\newtheorem*{defn*}{Definition}
\newtheorem{rem}[thm]{Remark}

\newtheorem*{assumption*}{Assumption}

\theoremstyle{remark}

\newcommand{\R}{\mathbb{R}}

\newcommand{\calU}{\mathcal{U}}
\newcommand{\calD}{\mathcal{D}}

\newcommand{\dd}{{\mathrm{d}}}

\newcommand{\C}{\mathbb{C}}

\newcommand{\Der}{{\mathrm{Der}}}

\newcommand{\T}{\textsf{T}}

\newcommand{\de}{\partial}

\def\gpd{\,\lower1pt\hbox{$\longrightarrow$}\hskip-.24in\raise2pt
               \hbox{$\longrightarrow$}\,}

\let\Hat=\widehat

\newcommand{\Sym}{\textnormal{Sym}}

\makeatletter
\newcommand{\upint}{\DOTSI\upintop\ilimits@}
\newcommand{\upoint}{\DOTSI\upointop\ilimits@}
\makeatother

\usetikzlibrary{positioning}
\usepackage{tikz-feynman} 
\tikzfeynmanset{compat=1.1.0}

\tikzfeynmanset{
  fermion1/.style={
    /tikz/postaction={
      /tikz/decoration={
        markings,
        mark=at position 0.5 with {
          \node[
            transform shape,
            xshift=-0.5mm,
            fill,
            dart tail angle=100,
            inner sep=1.3pt,
            draw=none,
            dart
          ] { };
        },
      },
      /tikz/decorate=true,
    },
  },
  fermion2/.style={
    /tikz/postaction={
      /tikz/decoration={
        markings,
        mark=at position 0.6 with {
          \arrow{>[length=6pt, width=5pt]};
        },
      },
      /tikz/decorate=true,
    },
  },
}

\usepackage{todonotes}

\makeatletter
\providecommand\@dotsep{5}
\renewcommand{\listoftodos}[1][\@todonotes@todolistname]{%
  \@starttoc{tdo}{#1}}
\makeatother

\makeatletter

\pgfdeclareshape{genus}{
 \anchor{center}{\pgfpointorigin}
\backgroundpath{
    \begingroup
    \tikz@mode
    \iftikz@mode@fill

         \pgfpathmoveto{\pgfqpoint{-0.78cm}{-.17cm}}
     \pgfpathcurveto %
            {\pgfpoint{-0.35cm}{-.44cm}}
        {\pgfpoint{0.35cm}{-.44cm}}
        {\pgfpoint{.78cm}{-0.17cm}} 
     \pgfpathmoveto{\pgfqpoint{-0.78cm}{-0.17cm}}
     \pgfpathcurveto %
            {\pgfpoint{-0.25cm}{.25cm}}
        {\pgfpoint{.25cm}{.25cm}}
        {\pgfpoint{0.78cm}{-0.17cm}}
        \pgfusepath{fill}
        \fi

        \iftikz@mode@draw
     \pgfpathmoveto{\pgfqpoint{-1cm}{0cm}}
     \pgfpathcurveto %
            {\pgfpoint{-0.5cm}{-.5cm}}
        {\pgfpoint{0.5cm}{-.5cm}}
        {\pgfpoint{1cm}{0cm}}

     \pgfpathmoveto{\pgfqpoint{-0.75cm}{-0.15cm}}
     \pgfpathcurveto %
            {\pgfpoint{-0.25cm}{.25cm}}
        {\pgfpoint{.25cm}{.25cm}}
        {\pgfpoint{0.75cm}{-0.15cm}}
              \pgfusepath{stroke}
        \fi
        \endgroup
    }
    }

    \makeatother

\title[Computation of Kontsevich Weights]{Computation of Kontsevich Weights of Connection and Curvature Graphs for Symplectic Poisson Structures}
\author[N. Moshayedi]{Nima Moshayedi}
\address{Institut f\"ur Mathematik\\ Universit\"at Z\"urich\\ 
Winterthurerstrasse 190
CH-8057 Z\"urich}
\email[N.~Moshayedi]{nima.moshayedi@math.uzh.ch}
\author[F. Musio]{Fabio Musio}
\address{Institut f\"ur Mathematik\\ Universit\"at Z\"urich\\ 
Winterthurerstrasse 190
CH-8057 Z\"urich}
\email[F.~Musio]{fabio.musio@uzh.ch}
\thanks{This research was (partly) supported by the NCCR SwissMAP, funded by the Swiss National Science
Foundation, and by the SNF grant No. 200020\_172498/1.}

\begin{document}

\maketitle


\tikzset{residual/.style={draw, shape=circle, black,inner sep=1pt}}

\begin{abstract}
    We give a detailed explicit computation of weights of Kontsevich graphs which arise from connection and curvature terms within the globalization picture as in \cite{CMW3} for the special case of symplectic manifolds. We will show how the weights for the curvature graphs can be explicitly expressed in terms of the hypergeometric function as well as by a much simpler formula combining it with the explicit expression for the weights of its underlined connection graphs. Moreover, we consider the case of a cotangent bundle, which will simplify the curvature expression significantly.
\end{abstract}

\tableofcontents

\section{Introduction}
\label{sec:Introduction}
\subsection{Motivation}
In \cite{K} Kontsevich proved that the differential graded Lie algebra (DGLA)\footnote{Endowed with the zero differential and the Schouten--Nijenhuis bracket.} of multivector fields on an open subset $M\subset \R^d$ is $L_\infty$-quasi-isomorphic to the DGLA\footnote{Endowed with the Hochschild differential and the Gerstenhaber bracket.} of multidifferential operators on functions on $M$, i.e. there exists an $L_\infty$-quasi-isomorphism
\begin{equation}
    \calU\colon T_{poly}(M)\to D_{poly}(M),
\end{equation}
such that its zeroth Taylor component $\calU^{(0)}$ is given by the Hochschild--Kostant--Rosenberg map. 
This result is known as the \emph{formality} theorem. If one restricts to the case of bivector fields and bidifferential operators, one can recover deformation quantization for Poisson manifolds. The resulting star product was also constructed in \cite{K} by an explicit formula.
In \cite{CFT,CF3,CMW3} a globalization picture was presented for this star product on any Poisson manifold $M$, including the construction of the local star product by using techniques of field theory, in particular the \emph{Poisson Sigma Model} \cite{I,SS1,CF1}. In \cite{CMW3} this construction was extended to manifolds with boundary as in the \emph{BV-BFV formalism} \cite{CMR1,CMR2,CattMosh1} which is a perturbative quantum gauge formalism compatible with cutting and gluing. A similar approach, as the one presented by Fedosov in \cite{Fedosov1994} for symplectic manifolds, was used, by considering notions of formal geometry. In particular, one starts with a \emph{formal exponential map} $\phi$ on the manifold $M$ and constructs a \emph{flat} connection $D_G$, called the \emph{classical Grothendieck connection}, on the completed symmetric algebra of the cotangent bundle $\Hat{\Sym}(T^*M)$. This construction can be deformed to the \emph{Weyl bundle} $\Hat{\Sym}(T^*M)[[\hbar]]$ and, as it was shown in \cite{CF3,CattaneoFelderTomassini2002,CMW3}, it induces a similar equation as the one in Fedosov's construction. In \cite{CMW3} it was shown that the different terms of this equation are given by a certain class of graphs. We want to give an explicit computation of the weights for these graphs.
Let $G_{n,m}$ denote the set of all \emph{admissible graphs} as in \cite{K} with $n$ vertices in the bulk of the upper half-plane $\mathbb{H}:=\{z\in\mathbb{C}|\;\text{Im}(z)>0\}$ and $m$ vertices on $\R$. Define a map 
\begin{equation}
    \calU_\Gamma\colon \bigwedge^nT_{poly}(M)\to D_{poly}(M)[1-n]
\end{equation}
using the $L_\infty$-morphism $\calU$.
Let $\pi$ be a Poisson structure on $\mathbb{R}^d$ and let $\xi,\zeta$ be any two vector fields on $\mathbb{R}^d$. Let us define 
\begin{align} 
P(\pi)&:=\sum\limits_{n\geq 0}\,\sum\limits_{\Gamma\in G_{n,2}}\frac{\hbar^n}{n!} w_\Gamma \mathcal{U}_{\Gamma}(\pi\wedge\cdots\wedge\pi), \label{def:P} \\
A(\xi,\pi)&:= \sum\limits_{n\geq 0}\,\sum\limits_{\Gamma\in G_{n+1,1}}\frac{\hbar^n}{n!} w_\Gamma \mathcal{U}_{\Gamma}(\xi\wedge\pi\wedge\cdots\wedge\pi), \label{def:A} \\
F(\xi,\zeta,\pi)&:= \sum\limits_{n\geq 0}\,\sum\limits_{\Gamma\in G_{n+2,0}}\frac{\hbar^n}{n!} w_\Gamma \mathcal{U}_{\Gamma}(\xi\wedge\zeta\wedge\pi\wedge\cdots\wedge\pi), \label{def:F}
\end{align}
where $w_\Gamma\in \R$ denotes the Kontsevich weight of the graph $\Gamma$. The term \eqref{def:P} represents Kontsevich's star product, \eqref{def:A} represents the \emph{deformed Grothendieck connection} $\calD_G:=D_G+O(\hbar)$ (see construction below) and \eqref{def:F} its curvature. 
Let us emphasize a bit more on the formal geometry construction. 
\subsection{Notions of Formal Geometry}
\label{subsec:formal_geometry}
Let $M$ be a smooth manifold and let $\phi \colon U \to M$ be a map where $U \subset TM$ is an open neighbourhood of the zero section. For $x \in M, y \in T_xM \cap U$ we write $\phi_x(y):=\phi(x,y)$. We say that $\phi$ is a \emph{generalized exponential map} if for all $x \in M$ we have that $\phi_x(0) = x$, and $\dd\phi_x(0) = \mathrm{id}_{T_xM}$. In local coordinates we can write 
\begin{equation}
\phi_x^{i}(y)=x^{i}+y^{i}+\frac{1}{2}\phi_{x,jk}^{i}y^jy^k+\frac{1}{3!}\phi^{i}_{x,jk\ell}y^jy^ky^\ell+\dotsm
\end{equation}
where the $x^i$ are coordinates on the base and the $y^i$ are coordinates on the fibers. 
We identify two generalized exponential maps if their jets agree to all orders. A \emph{formal exponential map} is an equivalence class of generalized exponential maps. It is completely specified by the sequence of functions $\left(\phi^i_{x,i_1\ldots i_k}\right)_{k=0}^{\infty}$. By abuse of notation, we will denote equivalence classes and their representatives by $\phi$. From a formal exponential map $\phi$ and a function $f \in C^{\infty}(M)$, we can produce a section $\sigma \in \Gamma(\Hat{\Sym}(T^*M))$ by defining $\sigma_x = \T\phi_x^*f$, where $\T$ denotes the Taylor expansion in the fiber coordinates around $y=0$ and we use any representative of $\phi$ to define the pullback. We denote this section by $\T\phi^*f$, it is independent of the choice of representative, since it only depends on the jets of the representative. 

As it was shown \cite{CF3,BCM,CMW4}, one can define a \emph{flat} connection $D_G$ on $\Hat{\Sym}(T^*M)$ with the property that $D_G\sigma = 0$ if and only if $\sigma = \T\phi^*f$ for some $f\in C^\infty(M)$.  
As already mentioned before, this connection is called the \emph{classical Grothendieck connection}.
In fact, $D_G = \dd_x + L_R$ where $R \in\Omega^1(M,\Der(\Hat{\Sym}(T^*M)))$ is a 1-form with values in derivations of $\Hat{\Sym}(T^*M)$, which we identify with $\Gamma(TM \otimes \Hat{\Sym}(T^*M))$. We have denoted by $\dd_x$ the de Rham differential on $M$ and by $L$ the Lie derivative. 
In coordinates we have 
\begin{equation}
    R(\sigma)_\ell=-\frac{\de\sigma}{\de y^j}\left(\left(\frac{\de\phi}{\de y}\right)^{-1}\right)_k^j\frac{\de\phi^k}{\de x^\ell}.
\end{equation}
Define $R(x,y):=R_\ell(x,y)\dd x^\ell$, $R_\ell(x,y):=R_\ell^j(x,y)\frac{\de}{\de y^j}$, $R^j(x,y):=R^j_\ell(x,y)\dd x^\ell$, and 
\begin{equation}
    R_\ell^j=-\left(\left(\frac{\de\phi}{\de y}\right)^{-1}\right)_k^j\frac{\de\phi^k}{\de x^\ell}=-\delta_\ell^j+O(y).
\end{equation}


For $\sigma \in \Gamma(\Hat{\Sym}(T^*M))$, $L_R\sigma$ is given by the Taylor expansion (in the $y$ coordinates) of $$-\dd_y\sigma \circ (\dd_y\phi)^{-1} \circ \dd_x\phi \colon \Gamma(TM) \to \Gamma(\Hat{\Sym}(T^*M)),$$
where we denote by $\dd_y$ the de Rham differential on the fiber.
This shows that $R$ does not depend on the choice of coordinates. One can generalize this also for any fixed vector $\xi = \xi^{i}(x) \frac{\partial}{\partial x^i}\in T_xM$, instead of just considering the de Rham differential $\dd_x$, by 
\begin{equation} 
\label{eq:Aconnection_G}
D_G^{\xi} = \xi + \Hat{\xi},
\end{equation}
where 
\begin{equation} 
\label{eq:formal_vector}
\Hat{\xi}(x,y) = \iota_{\xi}R(x,y) = \xi^i(x)R_\ell^j(x,y)\frac{\partial}{\partial y^j}. 
\end{equation}
Here $\xi^{i}(x)$ would replace the 1-form part $\dd x^{i}$. 

\vspace{0.5cm}

This paper is based on the master thesis \cite{Musio}.

\subsection*{Acknowledgements} 
We would like to thank Alberto Cattaneo and Konstantin Wernli for useful discussions.

\section{Computation of Kontsevich Weights} \label{sec:Computation Kontsevich}
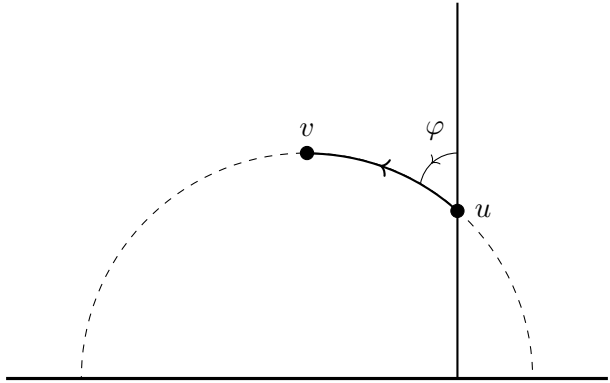
\begin{figure}[ht] 
\centering
\begin{tikzpicture}
\node[vertex, label=right: $u$] (u) at (2,2.23){};
\node[vertex, label=above: $v$] (v) at (0,3){};
\draw[very thick] (-4,0) -- (4,0);
\draw[thick] (2,0) -- (2,5);
\draw[dashed] (-3,0) -- (3,0) arc(0:180:3);
\draw[thick, fermion] (u) arc (48.18:90:3);
\draw[fermion] (2,3) arc (90:170:0.5);
\node[] (phi) at (1.7,3.3){$\varphi$};
\end{tikzpicture}
\caption{Illustartion of the angle function $\varphi$.}
\label{fig:angle function}
\end{figure}

Let $(M,\omega)$ be a symplectic manifold regarded as a Poisson manifold with Poisson structure $\pi$ induced by the symplectic form $\omega$. Moreover let $\phi\colon TM\supset U\rightarrow M$ be a formal exponential map and denote by $\mathsf{T}$ the Taylor expansion in fiber coordinates around zero. Anticipating the computation of the star product $P(\mathsf{T}\phi^*\pi)$, the connection $1$-form $A(R,\mathsf{T}\phi^*\pi)$ and its curvature $2$-form $F(R,R,\mathsf{T}\phi^*\pi)$ as in \cite{CMW3}, we will explicitly compute the Kontsevich weights of three families of graphs in this section.  
Throughout the paper we use the harmonic angle function 
\begin{align} \label{angle function}
\varphi(u,v)=\text{arg}\left(\frac{v-u}{v-\bar{u}}\right)=\frac{1}{2i}\text{log}\left(\frac{(v-u)(\bar{v}-u)}{(v-\bar{u})(\bar{v}-\bar{u})}\right)
\end{align}
which measures the angle on $\mathbb{H}\cup\mathbb{R}$ as depicted in Figure \ref{fig:angle function}.

The propagator used in the computation of the Kontsevich weights is then simply given by $\dd\varphi(u,v)$ and is usually called the \emph{Kontsevich propagator}. Now let $\Gamma\in G_{n,m}$ be an admissible graph with $n$ vertices of first type, $m$ vertices of second type and $2n+m-2$ edges. We use this propagator to compute the Kontsevich weight \cite{K} $w_{\Gamma}$ of $\Gamma$ as 
\begin{align} \label{Kontsevich weight}
w_\Gamma=\int\limits_{\bar C_{n,m}}\omega_\Gamma.
\end{align}
Here $\bar C_{n,m}$ denotes the \emph{Fulton--MacPherson/Axelrod--Singer (FMAS) compactification} \cite{FulMacPh,AS2} of the configuration space $C_{n,m}$ of $n$ points in $\mathbb{H}$ and $m$ points on $\R$ modulo scaling and translation.
Let us briefly recall the construction of the needed configuration spaces. We define the open configuration space
\begin{equation}
\label{eq:conf}
    \mathrm{Conf}_{n,m}=\{(x_1,\ldots,x_n,q_1,\ldots,q_m)\in \mathbb{H}^n\times \R^m\mid x_i\not=x_j,\,\forall i\not=j,\,q_1<\ldots<q_m\}.
\end{equation}
The 2-dimensional real Lie group of orientation preserving affine transformations of the real line
\begin{equation}
    G^{(1)}=\{z\mapsto az+b\mid a,b\in \R,\, a>0\}
\end{equation}
acts freely on $\mathrm{Conf}_{n,m}$. One can check that the quotient space $C_{n,m}:=\mathrm{Conf}_{n,m}/G^{(1)}$ is in fact a smooth manifold of dimension $2n+m-2$.
The differential form $\omega_\Gamma$ on $\bar C_{n,m}$ is given by
\begin{align}
\omega_\Gamma=\frac{1}{(2\pi)^{2n+m-2}}\bigwedge\limits_{\text{edges $e$}}\dd\varphi_e,
\end{align}
where the wedge product is over all $2n+m-2$ edges $e$ of the graph $\Gamma$.
Let $n\geq 2$ and define 
\begin{equation}
    \mathrm{Conf}_n=\{(x_1,\ldots,x_n)\in \mathbb{C}^n\mid x_i\not=x_j,\,\forall i\not=j\}.
\end{equation}
We have an action on $\mathrm{Conf}_n$ by the 3-dimensional Lie group 
\begin{equation}
    G^{(2)}=\{z\mapsto az+b\mid a\in \R,\,b\in \mathbb{C},\,a>0\}.
\end{equation}
Again, one can check that the quotient space $C_n:=\mathrm{Conf}_n/G^{(2)}$ is a smooth manifold of dimension $2n-3$. Also here, we will denote its FMAS compactification by $\bar C_n$. We refer to \cite{K} for a more detailed construction.
To simplify the notation we will use graphical language, where the figure below
corresponds to a factor of $\dd(\varphi(u,v)^n)/(2\pi)^n$ in $w_\Gamma$. If there is no $n$ above the arrow it simply means that $n=1$.

\begin{figure}[ht] 
 \centering
 \includegraphics[width=0.2\textwidth]{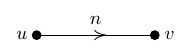}
 \label{fig:graphical language}
\end{figure}

We know that the dimension of the configuration space $C_{n,m}$ is $2n+m-2$, and since we work on a symplectic manifold $M$ (with Darboux coordinates around each point $x\in M$), a vertex of first type is either a vertex representing the tensor $\mathsf{T}\phi^{*}_{x}\pi$, which we will call a $\mathsf{T}\phi^{*}_{x}\pi$-vertex, with precisely two outgoing and no incoming edges, or a vertex representing the 1-form $R$, which we will call an $R$-vertex, with precisely one outgoing edge and arbitrarily many incoming edges \cite{CMW3,Mosh1}. So we may write $n=p+r$, where $p$ is the number of $\mathsf{T}\phi^{*}_{x}\pi$-vertices and $r$ is the number of $R$-vertices. We then have that $\text{deg}(\omega_\Gamma)=2p+r$, and in order for the integral (\ref{Kontsevich weight}) not to vanish, we must have that $\omega_\Gamma$ is a top form, i.e. that $2n+m-2=2p+r$. This then implies that
\begin{align}
r+m=2
\end{align}
So we have to distinguish three different cases, namely $(r,m)=(2,0)$, $(r,m)=(1,1)$ and $(r,m)=(0,2)$, which we will treat separately in what follows.

\begin{rem}
Actually, we will see below that all the integrals over the non-compactified configuration spaces $C_{n,m}$ of the graphs we are considering converge and are thus finite. So it is not necessary to work with the compactifications.  
\end{rem}

\subsection{Case 1: No Boundary Vertices} \label{subsec: no boundary vertices}
\begin{figure}[ht] 
 \centering
 \includegraphics[width=0.6\textwidth]{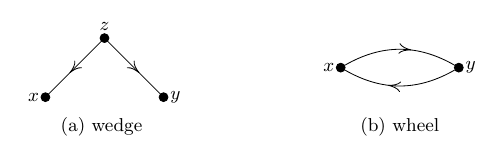}
 \caption{Graphs in the case $(r,m)=(2,0)$ consist of: (a) wedges stemming from $\mathsf{T}\phi^{*}_{x}\pi$-vertices attached to (b) a wheel stemming from the two $R$-vertices.}
 \label{fig:Kontsevich weight 1}
\end{figure}

We will first treat the case $(r,m)=(2,0)$, i.e. the case where we have no boundary vertices and exactely two $R$-vertices. In that case we get a family of graphs $(\Gamma_n)_{n\geq 0}$, where $\Gamma_n$ is the graph with $n$ wedges as in Figure \ref{fig:Kontsevich weight 1}(a) (stemming from $n$ $\mathsf{T}\phi^{*}_{x}\pi$-vertices) attached to the wheel as in Figure \ref{fig:Kontsevich weight 1}(b) (stemming from the two $R$-vertices).

\begin{figure}[ht] 
 \centering
 \includegraphics[width=0.8\textwidth]{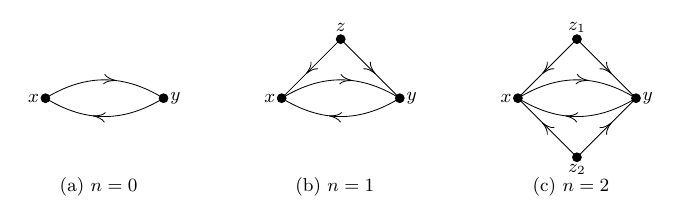}
 \caption{Graphs $\Gamma_n$ for (a) $n=0$, (b) $n=1$ and (c) $n=2$ wedges attached to the wheel.}
 \label{fig:Kontsevich weight 2}
\end{figure}

Examples of the graphs $\Gamma_n$ are given in Figure \ref{fig:Kontsevich weight 2} below for $n=0,1,2$.

The Kontsevich weight of the graph $\Gamma_n$ for $n\geq 0$ is given by
\begin{align} \label{Kontsevich weight family 1}
w_{\Gamma_{n}}=\frac{1}{(2\pi)^{2n+2}}\int\limits_{C_{n+2,0}}\dd\varphi(x,y)\dd\varphi(y,x)\dd\varphi(z_1,x)\dd\varphi(z_1,y)\cdots \dd\varphi(z_n,x)\dd\varphi(z_n,y).
\end{align}

\begin{rem}
We will omit the wedge product if it is clear. Moreover,
for $n=0$ we simply set $\dd\varphi(z_1,x)\dd\varphi(z_1,y)\cdots \dd\varphi(z_n,x)\dd\varphi(z_n,y)=1$ in the integral above.
\end{rem}

\begin{rem}
The sign of the weight $\omega_\Gamma$ depends on the ordering of the edges of the graph $\Gamma$ (i.e. the ordering of the propagator $1$-forms in the integrand), and thus the ordering must always be specified. Throughout this whole section, we will stick to the ordering given in \eqref{Kontsevich weight family 1}.
\end{rem}

The goal now is to compute \eqref{Kontsevich weight family 1} explicitly. We will do this in several steps, mainly using Stokes' theorem as in \cite{VanBergh2009}.

\subsubsection{Step 1} \label{subsubsec step 1 family 1}
In a first step, we want to integrate out the wedges. More precisely, for a wedge as in Figure \ref{fig:Kontsevich weight 1}(a) we want to compute the corresponding integral 
\begin{align} \label{int step 1}
\frac{1}{(2\pi)^2}\int\limits_{z\in\mathbb{H}\backslash\{x,y\}}\dd\varphi(z,x)\dd\varphi(z,y),
\end{align}
i.e. we want to integrate out $z$ (with $x,y\in\mathbb{H}$ fixed). To do this we make a branch cut such that $\varphi(z,x)\in(0,2\pi)$ and use Stokes' theorem
\begin{align} 
\int\limits_{z\in\mathbb{H}\backslash\{x,y\}}\dd\varphi(z,x)\dd\varphi(z,y)=\int\limits_{\partial}\varphi(z,x)\dd\varphi(z,y),
\end{align}
where $\partial$ is the boundary of the integration domain depicted in Figure \ref{fig:Kontsevich integral 1} below. 

\begin{figure}[ht] 
 \centering
 \includegraphics[width=1.0\textwidth]{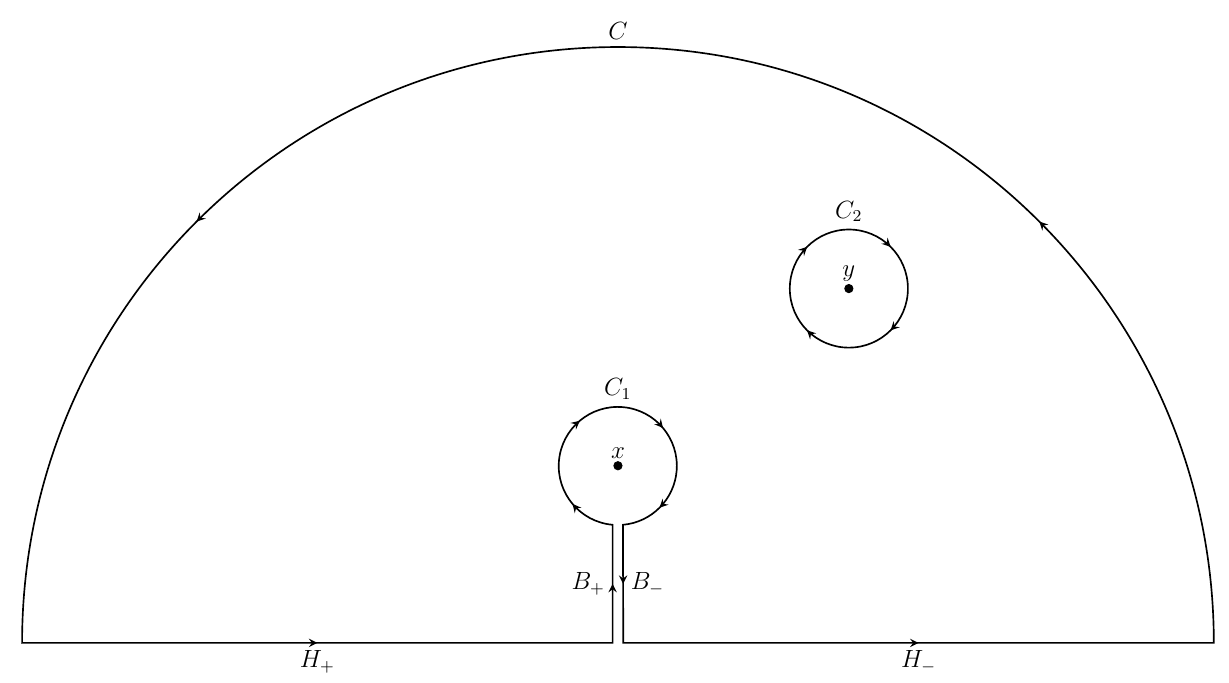}
 \caption{Boundary $\partial$ of the integration domain: $C$ is the half-circle at infinity, $B_{+}$ and $B_{-}$ are infinitesimally close together, the circles $C_1$ and $C_2$ have infinitesimal radius and $H_{+}\cup H_{-}$ is the real line.}
 \label{fig:Kontsevich integral 1}
\end{figure} 

Now using (\ref{angle function}) we can discuss the different boundary components:

\begin{itemize}
\item On $H_{+}\cup H_{-}$: $z\in\mathbb{R}$ and hence $\dd\varphi(z,y)=\dd\text{arg}(1)=0$
\item On $B_{+}$: $\varphi(z,x)= 2\pi$
\item On $B_{-}$: $\varphi(z,x)=0$ 
\item On $C_1$: $z=x+\varepsilon \mathrm{e}^{-i\theta}$ for $\varepsilon\rightarrow 0$ $\implies$ $\dd\varphi(z,y)=\dd\text{arg}\left(\frac{y-x}{y-\bar{x}}\right)=0$
\item On $C_2$: $z=y+\varepsilon \mathrm{e}^{-i\theta}$ for $\varepsilon\rightarrow 0$ and $\theta\in[0,2\pi)$ $\implies$ $\varphi(z,x)\rightarrow\varphi(y,x)$, $\dd\varphi(z,y)=-\dd\theta$
\item On $C$: $z=R\mathrm{e}^{i\theta}$ for $R\rightarrow\infty$ and $\theta\in[0,\pi]$ $\implies$ $\varphi(z,x)=\varphi(z,y)=2\theta$, $\dd\varphi(z,y)=2\dd\theta$
\end{itemize} 
We then finally get
\begin{align} \label{result wedges}
\begin{split}
\int\limits_{z\in\mathbb{H}\backslash\{x,y\}}\dd\varphi(z,x)\dd\varphi(z,y)&=\int\limits_{\partial}\varphi(z,x)\dd\varphi(z,y)=2\pi\int\limits_{B_{+}}\dd\varphi(z,y)+\int\limits_0^{\pi}4\theta \dd\theta-\varphi(y,x)\int\limits_0^{2\pi}\dd\theta
\\&=2\pi\big(\varphi(x,y)-\varphi(y,x)+[x;y]\pi\big),
\end{split}
\end{align}
where
\begin{align} \label{sign function}
[x;y]=
\begin{cases} 
+1, & \text{if }\mathrm{Re}(x)>\mathrm{Re}(y)\\ -1, & \text{if }\mathrm{Re}(x)<\mathrm{Re}(y)
\end{cases}
\end{align}
Dividing the result (\ref{result wedges}) by $(2\pi)^{2}$ (as in (\ref{int step 1})), we get
\begin{align} \label{result wedges 2}
\frac{1}{2\pi}\big(\varphi(x,y)-\varphi(y,x)\big)\pm\frac{1}{2},
\end{align}
which agrees with the result given in \cite[Lemma 3.3]{BanksPanzerPym2018}.

Finally, consider the limit $(x,y)\rightarrow (p,q)$ for $p,q\in\mathbb{R}$ with $p<q$. Using (\ref{result wedges 2}) and the fact that $\varphi(p,q)=2\pi$ and $\varphi(q,p)=0$, we can compute the Kontsevich weight of the graph below.

\begin{figure}[ht] 
 \centering
 \includegraphics[width=0.3\textwidth]{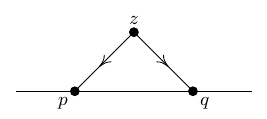}
\end{figure} 

We get that the integral of the form representing this graph over $C_{1,2}$ is equal to $\frac{1}{2}$, which agrees with the result in \cite[Section 6.4.3]{K}.


\subsubsection{Step 2}
In a second step we want to compute the weight of the graph

\begin{figure}[ht] 
 \centering
 \includegraphics[width=0.25\textwidth]{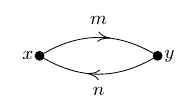}
\end{figure} 

where $n,m\geq 1$ and we use the notation introduced above, i.e. we want to explicitly compute the integral
\begin{align} \label{int step 2}
\frac{1}{(2\pi)^{n+m}}\int\limits_{C_{2,0}}\dd\varphi(x,y)^m \dd\varphi(y,x)^{n}.
\end{align}
As before we make the branch cut such that $\varphi(x,y)\in (0,2\pi)$ and use Stokes' theorem
\begin{align}
\int\limits_{y\in\mathbb{H}\backslash\{x\}}\dd\varphi(x,y)^m \dd\varphi(y,x)^{n}=\int\limits_{\partial}\varphi(x,y)^m \dd\varphi(y,x)^{n}
\end{align}
with boundary $\partial$ of the integration domain depicted in Figure \ref{fig:Kontsevich integral 2} below.

\begin{figure}[ht] 
 \centering
 \includegraphics[width=0.9\textwidth]{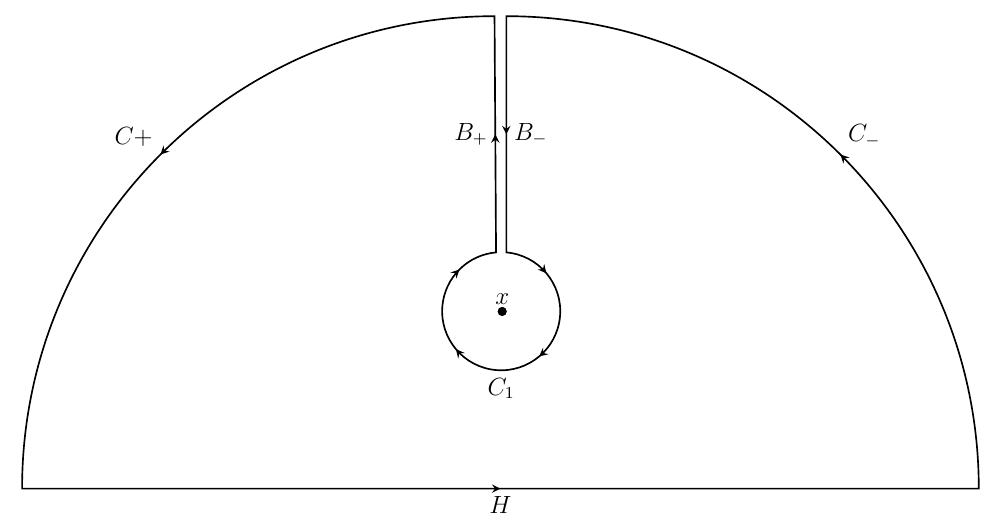}
 \caption{Boundary $\partial$ of the integration domain: $C_{-}\cup C_{+}$ is the half-circle at infinity, $B_{+}$ and $B_{-}$ are infinitesimally close together, the circle $C_1$ has infinitesimal radius and $H$ is the real line.}
 \label{fig:Kontsevich integral 2}
\end{figure} 

Again, we discuss the different boundary components:
\begin{itemize}
\item On $H$: $y\in\mathbb{R}$ $\implies$ $\dd\varphi(y,x)=\dd\text{arg}(1)=0$
\item On $B_{-}\cup B_{+}$: $\dd\varphi(y,x)=0$
\item On $C_{-}$: $y=R\mathrm{e}^{i\theta}$ for $R\rightarrow \infty$ $\implies$ $\varphi(x,y)= 2\pi$
\item On $C_{+}$: $y=R\mathrm{e}^{i\theta}$ for $R\rightarrow \infty$ $\implies$ $\varphi(x,y)= 0$
\item On $C_1$: $y=x+\varepsilon \mathrm{e}^{-i\theta}$ for $\varepsilon\rightarrow 0$ and $\theta\in(-\frac{\pi}{2},\frac{3\pi}{2})$ $\implies$ $\varphi(x,y)=\frac{3\pi}{2}-\theta$ and 
$\\\varphi(y,x)=
\begin{cases} 
\frac{\pi}{2}-\theta, & \text{for }\theta\in(-\frac{\pi}{2},\frac{\pi}{2}) \\ \frac{5\pi}{2}-\theta, & \text{for }\theta\in(\frac{\pi}{2},\frac{3\pi}{2})
\end{cases}$
\end{itemize} 
With this, we compute the integral
\begin{align}
\begin{split}
&\int\limits_{y\in\mathbb{H}\backslash\{x\}}\dd\varphi(x,y)^m \dd\varphi(y,x)^{n}=\int\limits_{\partial}\varphi(x,y)^m \dd\varphi(y,x)^{n}=(2\pi)^m\int\limits_{C_{-}}\dd\varphi(y,x)^n+\int\limits_{C_{1}}\varphi(x,y)^m \dd\varphi(y,x)^{n}
\\&=(2\pi)^m\pi^{n}-n\int\limits_{-\frac{\pi}{2}}^{\frac{\pi}{2}}\left(\frac{3\pi}{2}-\theta\right)^m\left(\frac{\pi}{2}-\theta\right)^{n-1}\dd\theta-n\int\limits_{\frac{\pi}{2}}^{\frac{3\pi}{2}}\left(\frac{3\pi}{2}-\theta\right)^m\left(\frac{5\pi}{2}-\theta\right)^{n-1}\dd\theta.
\end{split}
\end{align}
Now we use the substitution $a=\frac{\pi}{2}-\theta$ to compute
\begin{align} \label{substitution 1}
\begin{split}
\int\limits_{-\frac{\pi}{2}}^{\frac{\pi}{2}}\left(\frac{3\pi}{2}-\theta\right)^m\left(\frac{\pi}{2}-\theta\right)^{n-1}\dd\theta&=\int\limits_{0}^{\pi}(\pi+a)^m a^{n-1}\dd a=\sum\limits_{k=0}^{m}{m\choose k}\pi^{k}\int\limits_{0}^{\pi}a^{m+n-k-1}\dd a
\\&=\sum\limits_{k=0}^{m}{m\choose k}\frac{\pi^{m+n}}{m+n-k}.
\end{split}
\end{align}
Similarly, we use the substitution $a=\frac{3\pi}{2}-\theta$ to compute
\begin{align} \label{substitution 2}
\begin{split}
\int\limits_{\frac{\pi}{2}}^{\frac{3\pi}{2}}\left(\frac{3\pi}{2}-\theta\right)^m\left(\frac{5\pi}{2}-\theta\right)^{n-1}\dd\theta&=\int\limits_{0}^{\pi}a^m (\pi+a)^{n-1}\dd a=\sum\limits_{k=0}^{n-1}{n-1\choose k}\pi^k\int\limits_{0}^{\pi}a^{m+n-k-1}\dd a
\\&=\sum\limits_{k=0}^{n-1}{n-1\choose k}\frac{\pi^{m+n}}{m+n-k}.
\end{split}
\end{align}
Putting everything together we get
\begin{align} \label{result step 2}
\int\limits_{y\in\mathbb{H}\backslash\{x\}}\dd\varphi(x,y)^m \dd\varphi(y,x)^{n}=\left(2^m-\sum\limits_{k=0}^{m}{m\choose k}\frac{n}{m+n-k}-\sum\limits_{l=0}^{n-1}{n-1\choose l}\frac{n}{m+n-l}\right)\pi^{m+n}.
\end{align}
It is not hard to see that for $n=1$ the above formula simplifies to
\begin{align}
\int\limits_{y\in\mathbb{H}\backslash\{x\}}\dd\varphi(x,y)^m \dd\varphi(y,x)=2^{m}\left(1-\frac{2}{m+1}\right)\pi^{m+1},
\end{align}
which agrees with the result in \cite[Section 4]{VanBergh2009}. 

\subsubsection{Step 3}
In a third step we want to compute an integral similar to (\ref{int step 2}), but with an additional factor $[x;y]$ as defined in (\ref{sign function}). So we want to compute the integral
\begin{align} \label{int step 3}
\frac{1}{(2\pi)^{n+m}}\int\limits_{y\in\mathbb{H}\backslash\{x\}}[x;y]\dd\varphi(x,y)^m \dd\varphi(y,x)^{n}
\end{align}
As usual, we use Stokes' theorem
\begin{align} 
\begin{split}
\int\limits_{y\in\mathbb{H}\backslash\{x\}}[x;y]\dd\varphi(x,y)^m \dd\varphi(y,x)^{n}&=\int\limits_{\substack{y\in\mathbb{H}\backslash\{x\}\\\mathrm{Re}(y)<\mathrm{Re}(x)}}\dd\varphi(x,y)^m \dd\varphi(y,x)^{n}-\int\limits_{\substack{y\in\mathbb{H}\backslash\{x\}\\\mathrm{Re}(y)>\mathrm{Re}(x)}}\dd\varphi(x,y)^m \dd\varphi(y,x)^{n}
\\&=\int\limits_{\partial_+}\varphi(x,y)^m \dd\varphi(y,x)^{n}-\int\limits_{\partial_-}\varphi(x,y)^m \dd\varphi(y,x)^{n}
\end{split}
\end{align}
with boundaries $\partial_{+}$ and $\partial_{-}$ of the integration domain depicted in Figure \ref{fig:Kontsevich integral 3} below. 

\begin{figure}[ht] 
 \centering
 \includegraphics[width=0.9\textwidth]{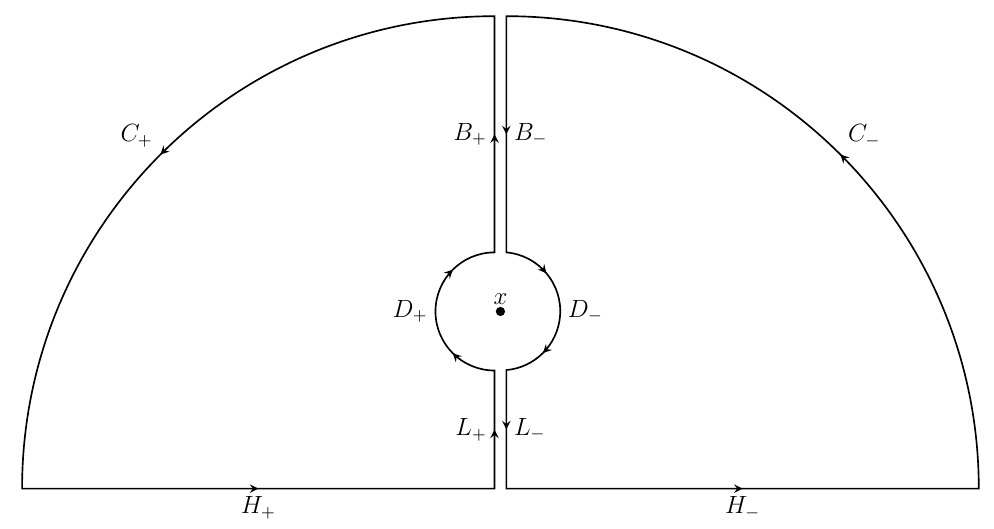}
 \caption{Boundaries $\partial_{+}$ (on the left) and $\partial_-$ (on the right) of the integration domain: $C_{-}\cup C_{+}$ is the half-circle at infinity, $B_{+}$ and $B_{-}$ as well as $L_+$ and $L_-$ are infinitesimally close together, the circle $D_+\cup D_-$ has infinitesimal radius and $H_+\cup H_-$ is the real line.}
 \label{fig:Kontsevich integral 3}
\end{figure} 
As before, we discuss the different boundary components:
\begin{itemize}
\item On $H_{+}\cup H_{-}$: $\dd\varphi(y,x)=0$
\item On $B_{\pm}\cup L_{\pm}$: $\dd\varphi(y,x)=0$
\item On $C_{-}$: $\varphi(x,y)=2\pi$
\item On $C_{+}$: $\varphi(x,y)=0$
\item On $D_{-}$: $y=x+\varepsilon \mathrm{e}^{-i\theta}$ for $\varepsilon\rightarrow 0$ and $\theta\in(-\frac{\pi}{2},\frac{\pi}{2})$ $\implies$ $\varphi(x,y)=\frac{3\pi}{2}-\theta$, $\varphi(y,x)=\frac{\pi}{2}-\theta$
\item On $D_{+}$: $y=x+\varepsilon \mathrm{e}^{-i\theta}$ for $\varepsilon\rightarrow 0$ and $\theta\in(\frac{\pi}{2},\frac{3\pi}{2})$ $\implies$ $\varphi(x,y)=\frac{3\pi}{2}-\theta$, $\varphi(y,x)=\frac{5\pi}{2}-\theta$
\end{itemize}
With this we compute the integral
\begin{align} \label{result step 3}
\begin{split}
&\int\limits_{y\in\mathbb{H}\backslash\{x\}}[x;y]\dd\varphi(x,y)^m \dd\varphi(y,x)^{n}=\int\limits_{\partial_+}\varphi(x,y)^m \dd\varphi(y,x)^{n}-\int\limits_{\partial_-}\varphi(x,y)^m \dd\varphi(y,x)^{n}
\\&=-n\int\limits_{\frac{\pi}{2}}^{\frac{3\pi}{2}}\left(\frac{3\pi}{2}-\theta\right)^{m}\left(\frac{5\pi}{2}-\theta\right)^{n-1}d\theta-(2\pi)^{m}\pi^{n}+n\int\limits_{-\frac{\pi}{2}}^{\frac{\pi}{2}}\left(\frac{3\pi}{2}-\theta\right)^{m}\left(\frac{\pi}{2}-\theta\right)^{n-1}d\theta
\\&=\left(-2^m+\sum\limits_{k=0}^{m}{m\choose k}\frac{n}{m+n-k}-\sum\limits_{l=0}^{n-1}{n-1\choose l}\frac{n}{m+n-l}\right)\pi^{m+n},
\end{split}
\end{align}
where we used (\ref{substitution 1}) and (\ref{substitution 2}) in the last step.

\subsubsection{Putting everything together}
Finally, we are able to compute the Kontsevich weight (\ref{Kontsevich weight family 1}) of the graphs $\Gamma_{n}$ described at the beginning of Section \ref{subsec: no boundary vertices}. Integrating over $z_{i}$ for $i=1,\ldots,n$, and applying the result for (\ref{result wedges}) obtained in the first step we get
\begin{align} \label{Kontsevich weight whole expression 1}
\begin{split}
w_{\Gamma_{n}}&=\frac{1}{(2\pi)^{2n+2}}\int\limits_{C_{n+2,0}}\dd\varphi(x,y)\dd\varphi(y,x)\dd\varphi(z_1,x)\dd\varphi(z_1,y)\cdots \dd\varphi(z_n,x)\dd\varphi(z_n,y)
\\&=\frac{1}{(2\pi)^{n+2}}\int\limits_{y\in\mathbb{H}\backslash\{x\}}\big(\varphi(x,y)-\varphi(y,x)+[x;y]\pi\big)^{n}\dd\varphi(x,y)\dd\varphi(y,x)
\\&=\frac{1}{(2\pi)^{n+2}}\sum\limits_{k=0}^{n}\sum\limits_{l=0}^{n-k}{n\choose k}{n-k\choose l}(-1)^{l}\int\limits_{y\in\mathbb{H}\backslash\{x\}}\varphi(x,y)^{n-k-l}\varphi(y,x)^ {l}([x;y]\pi)^{k}\dd\varphi(x,y)\dd\varphi(y,x)
\\&=\frac{1}{2^{n+2}}\sum\limits_{k=0}^{n}\sum\limits_{l=0}^{n-k}{n\choose k}{n-k\choose l}\frac{(-1)^{l}}{\pi^{n-k+2}(n-k-l+1)(l+1)}\int\limits_{y\in\mathbb{H}\backslash\{x\}}[x;y]^{k}\dd\varphi(x,y)^{n-k-l+1}\dd\varphi(y,x)^{l+1}.
\end{split}
\end{align} 
Note that for even $k$ we have 
\begin{align}
\begin{split}
&\int\limits_{y\in\mathbb{H}\backslash\{x\}}[x;y]^{k}\dd\varphi(x,y)^{n-k-l+1}\dd\varphi(y,x)^{l+1}=\int\limits_{y\in\mathbb{H}\backslash\{x\}}\dd\varphi(x,y)^{n-k-l+1}\dd\varphi(y,x)^{l+1}
\\&=\left(2^{n-k-l+1}-\sum\limits_{r=0}^{n-k-l+1}{n-k-l+1\choose r}\frac{l+1}{n-k-r+2}-\sum\limits_{s=0}^{l}{l\choose s}\frac{l+1}{n-k-s+2}\right)\pi^{n-k+2},
\end{split}
\end{align}
where we have used (\ref{result step 2}). Similarly, for odd $k$ we get
\begin{align}
\begin{split}
&\int\limits_{y\in\mathbb{H}\backslash\{x\}}[x;y]^{k}\dd\varphi(x,y)^{n-k-l+1}\dd\varphi(y,x)^{l+1}=\int\limits_{y\in\mathbb{H}\backslash\{x\}}[x;y]\dd\varphi(x,y)^{n-k-l+1}\dd\varphi(y,x)^{l+1}
\\&=\left(-2^{n-k-l+1}+\sum\limits_{r=0}^{n-k-l+1}{n-k-l+1\choose r}\frac{l+1}{n-k-r+2}-\sum\limits_{s=0}^{l}{l\choose s}\frac{l+1}{n-k-s+2}\right)\pi^{n-k+2},
\end{split}
\end{align}
where we have used (\ref{result step 3}).

We will now simplify the expressions we got. We start by observing a few things: 

First of all, we clearly have that
\begin{align} \label{n even 1}
\dd\varphi(x,y)^{n-k-l+1}\dd\varphi(y,x)^{l+1}=-\dd\varphi(y,x)^{l+1}\dd\varphi(x,y)^{n-k-l+1}.
\end{align}
Similarly, we also have that
\begin{align} \label{n even 2}
[x;y]=-[y;x].
\end{align}
Furthermore, we can obviously swap $x$ and $y$ in the integral and get the same result, i.e.
\begin{align} \label{n even 3}
\int\limits_{y\in\mathbb{H}\backslash\{x\}}[x;y]^{m}\dd\varphi(x,y)^{n-k-l+1}\dd\varphi(y,x)^{l+1}=\int\limits_{x\in\mathbb{H}\backslash\{y\}}[y;x]^{m}\dd\varphi(y,x)^{n-k-l+1}\dd\varphi(x,y)^{l+1}.
\end{align}
Now assume that $n$ is even. Applying (\ref{n even 1}), (\ref{n even 2}) and (\ref{n even 3}) to the last line of (\ref{Kontsevich weight whole expression 1}), it follows that
\begin{align} \label{Kontsevich weight n even 1}
w_{\Gamma_n}=\frac{1}{2^{n+2}}\sum\limits_{\substack{k=0\\k \text{ even}}}^{n}{n\choose k}{n-k\choose \frac{n-k}{2}}\frac{(-1)^{\frac{n-k}{2}}}{\pi^{n-k+2}\left(\frac{n-k}{2}+1\right)^2}\int\limits_{y\in\mathbb{H}\backslash\{x\}}\dd\varphi(x,y)^{\frac{n-k}{2}+1}\dd\varphi(y,x)^{\frac{n-k}{2}+1}.
\end{align}
So most of the terms cancel for $n$ even. 
\\Now using (\ref{result step 2}) we observe that
\begin{align} 
\begin{split}
&\frac{1}{\pi^{2m}}\int\limits_{y\in\mathbb{H}\backslash\{x\}}\dd\varphi(x,y)^m \dd\varphi(y,x)^{m}=2^m-\sum\limits_{k=0}^{m}{m\choose k}\frac{m}{2m-k}-\sum\limits_{l=0}^{m-1}{m-1\choose l}\frac{m}{2m-l}
\\&=2^m-\sum\limits_{k=0}^{m-1}\left({m\choose k}+{m-1\choose k}\right)\frac{m}{2m-k}-1=2^m-\sum\limits_{k=0}^{m-1}{m\choose k}-1=2^m-\sum\limits_{k=0}^{m}{m\choose k}=0.
\end{split}
\end{align}
Plugging this result into (\ref{Kontsevich weight n even 1}) with $m=\frac{n-k}{2}$ we finally get that
\begin{align} \label{Kontsevich weight n even 2}
w_{\Gamma_n}=0\qquad\text{for even }n\geq 0.
\end{align}
For $n$ odd the different terms in the last line of (\ref{Kontsevich weight whole expression 1}) do not cancel anymore. Instead, we will try to write (\ref{result step 2}) and (\ref{result step 3}) more compactly. To do this, let us introduce the so-called \emph{hypergeometric function} $_2 F_1(a,b;c;z)$. It is defined by the series
\begin{align}
_2 F_1(a,b;z;c):=\sum\limits_{k=0}^{\infty}\frac{(a)_k(b)_k}{(c)_k}\frac{z^k}{k!}
\end{align}
for $z\in\mathbb{C}$ with $|z|<1$, where $(a)_k$ is the \emph{Pochhammer symbol} given by
\begin{align}
(a)_k=
\begin{cases} 
1, & \text{if }k=0\\ a(a+1)\cdots(a+k-1), & \text{if }k>0
\end{cases}
\end{align}
It is not hard to see that the series terminates if either $a$ or $b$ is a non-positive integer. In that case the hypergeometric function reduces to a polynomial and can therefore also be defined for $|z|\geq 1$.
\\We can now use the hypergeometric function to write
\begin{align}
\sum\limits_{k=0}^{m}{m\choose k}\frac{1}{m+n-k}=\frac{_2 F_1(-m,-m-n;1-m-n;-1)}{m+n},
\end{align}
and 
\begin{align}
\sum\limits_{k=0}^{n-1}{n-1\choose k}\frac{1}{m+n-k}=\frac{_2 F_1(1-n,-m-n;1-m-n;-1)}{m+n}.
\end{align}
This allows us to write (\ref{result step 2}) as
\begin{align} \label{result step 2 simplified}
\begin{split}
&\int\limits_{y\in\mathbb{H}\backslash\{x\}}\dd\varphi(x,y)^m \dd\varphi(y,x)^{n}
\\&=\Big(2^{m}-\frac{n}{m+n}\big({_2 F_1}(-m,-m-n;1-m-n;-1)+{_2 F_1}(1-n,-m-n;1-m-n;-1)\big)\Big)\pi^{m+n},
\end{split}
\end{align}
and (\ref{result step 3}) as
\begin{align} \label{result step 3 simplified}
\begin{split}
&\int\limits_{y\in\mathbb{H}\backslash\{x\}}[x;y]\dd\varphi(x,y)^m \dd\varphi(y,x)^{n}
\\&=\left(-2^{m}+\frac{n}{m+n}\big({_2 F_1}(-m,-m-n;1-m-n;-1)-{_2 F_1}(1-n,-m-n;1-m-n;-1)\big)\right)\pi^{m+n}.
\end{split}
\end{align}
Plugging those results into (\ref{Kontsevich weight whole expression 1}) we finally get for all $n\geq 0$
\begin{align} \label{Kontsevich weight whole expression 2}
\begin{split}
w_{\Gamma_n}&=\frac{1}{2^{n+2}}\sum\limits_{k=0}^{n}\sum\limits_{l=0}^{n-k}{n\choose k}{n-k\choose l}\frac{(-1)^{l}}{(n-k-l+1)(l+1)}\Big((-1)^k 2^{n-k-l+1}
\\&-\frac{l+1}{n-k+2}\big({_2 F_1}(-l,-n+k-2;-n+k-1;-1)
\\&+(-1)^{k}{_2 F_1}(-n+k+l-1,-n+k-2;-n+k-1;-1)\big)\Big).
\end{split}
\end{align}
The Kontsevich weights of the first few graphs are given in Table \ref{table: Kontsevich weights 1} below.

\begin{table}[ht] 
\begin{center}
 \begin{tabular}{|c|cccccccccc|} 
 \hline
 $n$ & $0$ & $1$ & $2$ & $3$ & $4$ & $5$ & $6$ & $7$ & $8$ & $9$ \\ 
 \hline
 \\[-0.35cm]
$w_{\Gamma_n}$ & $0$ & $\frac{1}{24}$ & $0$ & $\frac{1}{320}$ & $0$ & $\frac{1}{2688}$ & $0$ & $\frac{1}{18432}$ & $0$ & $\frac{1}{112640}$  \\[0.08cm] 
\hline 
\end{tabular}
\end{center}
\caption{Kontsevich weights of the graphs $\Gamma_n$ for $n=0,1,\ldots,9$}
\label{table: Kontsevich weights 1}
\end{table}

As a sanity check we have the following: For $n=0$ the graph $\Gamma_0$ is just a wheel with two vertices (see Figure \ref{fig:Kontsevich weight 2}(a)) and its weight is zero according to \cite[Lemma 7.3]{K}. For $n=1$ the graph $\Gamma_1$ is just a wheel with two spokes pointing outward (see Figure \ref{fig:Kontsevich weight 2}(b)) and its weight is $\frac{1}{24}$ according to \cite[Proposition 1.1]{VanBergh2009}. So at least for $n=0,1$ our formula (\ref{Kontsevich weight whole expression 2}) for the Kontsevich weights $w_{\Gamma_n}$ produces the correct values.

\subsection{Case 2: One Boundary Vertex} \label{subsec: one boundary vertex}
\begin{figure}[ht] 
 \centering
 \includegraphics[width=0.55\textwidth]{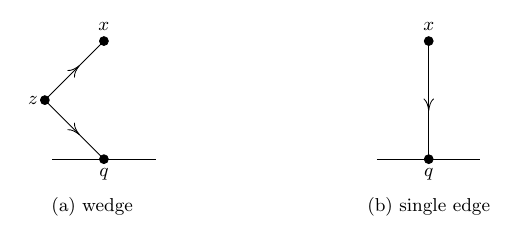}
 \caption{Graphs in the case $(r,m)=(1,1)$ consist of: (a) wedges stemming from $\mathsf{T}\phi^{*}_{x}\pi$-vertices attached to (b) a single edge from the $R$-vertex to the boundary vertex}
 \label{fig:Kontsevich weight 6}
\end{figure}

We will now treat the case $(r,m)=(1,1)$, i.e. the case where we have one boundary vertex and one $R$-vertex. In that case we get a family of graphs $(\Upsilon_n)_{n\geq 0}$, where $\Upsilon_n$ is the graph with $n$ wedges as in Figure \ref{fig:Kontsevich weight 6}(a) (stemming from $n$ $\mathsf{T}\phi_{x}^{*}\pi$-vertices) attached to the graph containing a single edge from the $R$-vertex to the boundary vertex as in Figure \ref{fig:Kontsevich weight 6}(b) below.

\begin{figure}[ht] 
 \centering
 \includegraphics[width=0.75\textwidth]{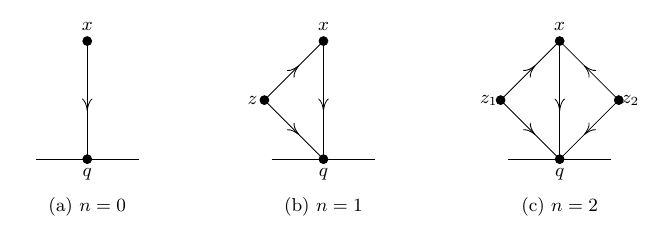}
 \caption{Graphs $\Upsilon_n$ for (a) $n=0$, (b) $n=1$ and (c) $n=2$ wedges attached to the single edge from the $R$-vertex to the boundary vertex}
 \label{fig:Kontsevich weight 7}
\end{figure}

Examples of the graphs $\Upsilon_n$ are given in Figure \ref{fig:Kontsevich weight 7} below for $n=0,1,2$.

The Kontsevich weight of the graph $\Upsilon_n$ for $n\geq 0$ is given by 
\begin{align} \label{Kontsevich weight family 2}
w_{\Upsilon_n}=\frac{1}{(2\pi)^{2n+1}}\int\limits_{C_{n+1,1}}\dd\varphi(x,q)\dd\varphi(z_1,x)\dd\varphi(z_1,q)\cdots \dd\varphi(z_n,x)\dd\varphi(z_n,q).
\end{align}

\begin{rem}
As before, the ordering of the edges of the graph $\Upsilon_n$ specified in (\ref{Kontsevich weight family 2}) above determines the sign of $w_{\Upsilon_n}$. Throughout this whole section we will stick to this ordering. 
\end{rem}

Again, the goal is to compute (\ref{Kontsevich weight family 2}) explicitly. As before, we will do this in several steps.

\subsubsection{Step 1}
For a wedge as in Figure \ref{fig:Kontsevich weight 6}(a) we want to compute the corresponding integral 
\begin{align}
\frac{1}{(2\pi)^2}\int\limits_{z\in\mathbb{H}\backslash\{x\}}\dd\varphi(z,x)\dd\varphi(z,q),
\end{align}
i.e. we want to integrate out $z$ (with $x,q\in\overline{\mathbb{H}}$ fixed). The computation is almost the same as the one we have already done in Section \ref{subsubsec step 1 family 1} above: Again we make a branch cut such that $\varphi(z,x)\in(0,2\pi)$ and use Stokes' theorem
\begin{align}
\int\limits_{z\in\mathbb{H}\backslash\{x\}}\dd\varphi(z,x)\dd\varphi(z,q)=\int\limits_{\partial}\varphi(z,x)\dd\varphi(z,q),
\end{align} 
where $\partial$ is the boundary of the integration domain depicted in Figure \ref{fig:Kontsevich integral 4} below.

\begin{figure}[ht] 
 \centering
 \includegraphics[width=1.0\textwidth]{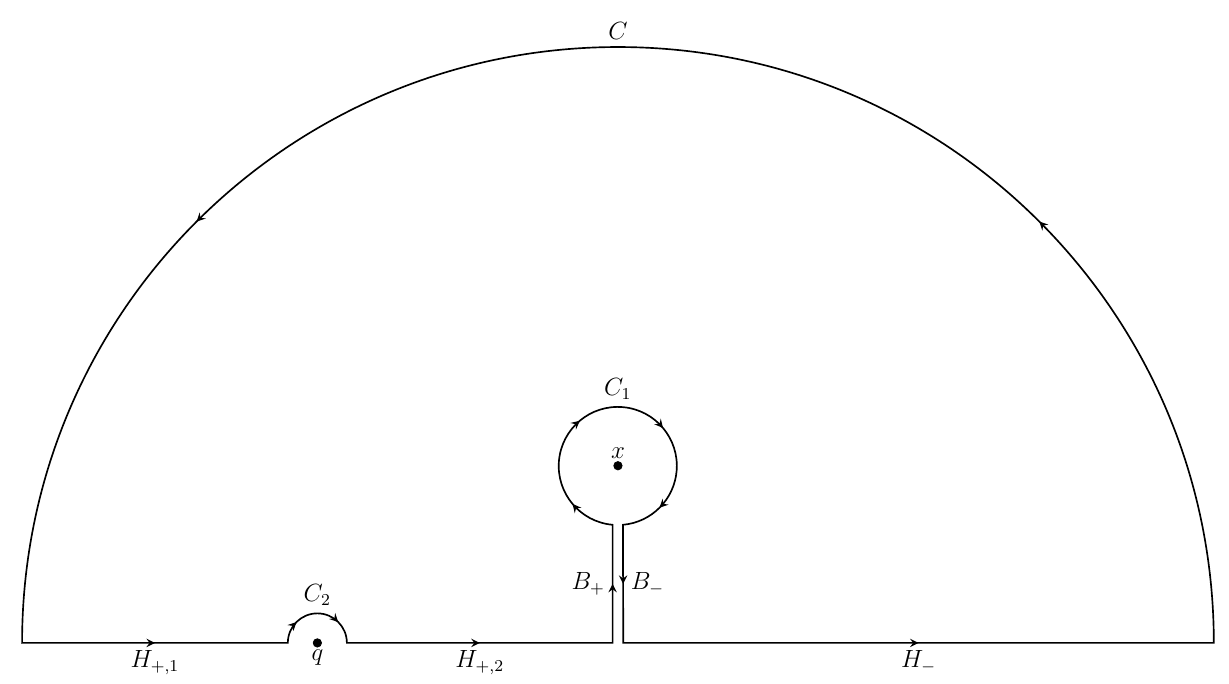}
 \caption{Boundary $\partial$ of the integration domain: $C$ is the half-circle at infinity, $B_{+}$ and $B_{-}$ are infinitesimally close together, the (half-)circles $C_1$ and $C_2$ have infinitesimal radius and $H_{+,1}\cup H_{+,2}\cup H_{-}$ is the real line.}
 \label{fig:Kontsevich integral 4}
\end{figure} 

Let us have a look at the different boundary components:
\begin{itemize}
\item On $H_{+,1}\cup H_{+,2}\cup H_{-}$: $z\in\mathbb{R}$ and hence $\dd\varphi(z,q)=0$
\item On $B_{+}$: $\varphi(z,x)= 2\pi$
\item On $B_{-}$: $\varphi(z,x)=0$ 
\item On $C_1$: $z=x+\varepsilon \mathrm{e}^{-i\theta}$ for $\varepsilon\rightarrow 0$ $\implies$ $\dd\varphi(z,q)=\dd\text{arg}\left(\frac{q-x}{q-\bar{x}}\right)=0$
\item On $C_2$: $z=q+\varepsilon \mathrm{e}^{-i\theta}$ for $\varepsilon\rightarrow 0$ and $\theta\in[-\pi,0]$ $\implies$ $\varphi(z,x)\rightarrow\varphi(q,x)$, $\varphi(z,q)=-2\theta$
\item On $C$: $z=R\mathrm{e}^{i\theta}$ for $R\rightarrow\infty$ and $\theta\in[0,\pi]$ $\implies$ $\varphi(z,x)=\varphi(z,q)=2\theta$
\end{itemize}
We can then compute the integral
\begin{align} \label{result boundary wedges}
\begin{split}
\int\limits_{z\in\mathbb{H}\backslash\{x,q\}}\dd\varphi(z,x)\dd\varphi(z,q)&=\int\limits_{\partial}\varphi(z,x)\dd\varphi(z,q)=2\pi\int\limits_{B_{+}}\dd\varphi(z,q)+\int\limits_0^{\pi} 4\theta \dd\theta-2\varphi(q,x)\int\limits_{-\pi}^{0}\dd\theta
\\&=2\pi\big(\varphi(x,q)-\varphi(q,x)+[x;q]\pi\big)
\end{split}
\end{align} 
where
\begin{align} \label{sign function 2}
[x;q]=
\begin{cases} 
+1, & \text{if }\mathrm{Re}(x)>q\\ -1, & \text{if }\mathrm{Re}(x)<q
\end{cases}
\end{align}
Dividing the result (\ref{result boundary wedges}) by $(2\pi)^2$ we get
\begin{align}
\frac{1}{2\pi}\big(\varphi(x,q)-\varphi(q,x)\big)\pm\frac{1}{2},
\end{align}
which agrees with the result in \cite[Lemma 5.3]{VanBergh2009}.
\\Finally, observe that one obtains (\ref{result boundary wedges}) by simply taking the limit $y\rightarrow q\in\mathbb{R}$ in (\ref{result wedges}).

\subsubsection{Step 2}
In a second step we want to compute the integral
\begin{align} \label{int boundary step 2}
\frac{1}{(2\pi)^{m+n}}\int\limits_{C_{1,1}}\varphi(q,x)^m \dd\varphi(x,q)^n
\end{align}
for $n\geq 1$ and $m\geq 0$.
\\First note that $C_{1,1}$, shown in Figure \ref{fig:config space 1} below, is a smooth manifold of dimension $1$ which is homeomorphic to an open interval and $\bar C_{1,1}$ is homeomorphic to a closed interval.

\begin{rem} 
We work with the standard orientation on $C_{1,1}$, which is induced by the standard orientation on the plane $\mathbb{R}^{2}$.
\end{rem}

\begin{figure}[ht] 
 \centering
 \includegraphics[width=0.5\textwidth]{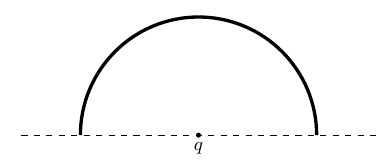}
 \caption{The manifold $C_{1,1}$ is the product of a (fixed) single point $q$ on the real line and an open half circle}
 \label{fig:config space 1}
\end{figure} 
It is not hard to see that the boundary $\partial C_{1,1}$ is just a two-element set. More precisely $\partial C_{1,1}=\{(q,s),(q,t)\}$ with $s<q$ and $t>q$ (for a more detailed treatment, see \cite{K,CMW3,CattMosh1}). 
\\But now we have to make a branch cut such that $\varphi(q,x)\in(0,2\pi)$. Then the boundary $\partial$ of the integration domain, depicted in Figure \ref{fig:config space 2} below, contains four points, namely 
\begin{align}
\partial=\{(q,s),(q,t),(q,y_+),(q,y_{-})\},
\end{align}
where $y$ is the point on the half circle directly above $q$, i.e. with Re$(y)=q$, and $y_{+}$ and $y_{-}$ are the limits $x\rightarrow y$ on the half circle from the left (i.e. from the region Re$(x)<q$ of the half-circle) and from the right (i.e. from the region Re$(x)>q$) respectively.

\begin{figure}[ht] 
 \centering
 \includegraphics[width=0.5\textwidth]{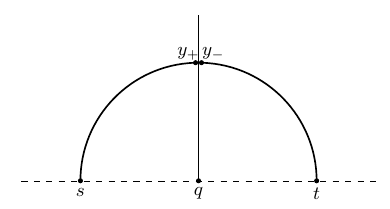}
 \caption{$C_{1,1}$ with branch cut and its boundary $\partial$ consisting of four points}
 \label{fig:config space 2}
\end{figure} 

Finally, using Stokes' theorem and the fact that $\dd\varphi(q,x)=0$ for $q\in\mathbb{R}$, we get
\begin{align} \label{result boundary step 2}
\begin{split}
&\int\limits_{C_{1,1}}\varphi(q,x)^m \dd\varphi(x,q)^n=\int\limits_{\partial }\varphi(q,x)^m \varphi(x,q)^n
\\&=\varphi(q,s)^m\varphi(s,q)^n-\varphi(q,y_+)^m\varphi(y_+,q)^n+\varphi(q,y_-)^m\varphi(y_-,q)^n-\varphi(q,t)^m\varphi(t,q)^n
\\&=
\begin{cases}
(2\pi)^n, & \text{if }m=0 \\ 2^m\pi^{m+n}, & \text{if } m>0
\end{cases}
\end{split}
\end{align}

\subsubsection{Step 3}
Let us start with writing (\ref{result boundary wedges}) differently as
\begin{align} \label{result boundary wedges new}
2\pi\big(\varphi(x,q)-\varphi(q,x)+\pi[x;q]\big)=2\pi\big(\varphi(x,q)-\varphi(q,x)-\pi+2\pi(x;q)\big)
\end{align}
where
\begin{align} \label{sign function new}
(x;q)=
\begin{cases} 
+1, & \text{if }\mathrm{Re}(x)>q\\ 0, & \text{if }\mathrm{Re}(x)<q
\end{cases}
\end{align}
In this step we then want to compute an integral similar to (\ref{int boundary step 2}), but with an additional factor $(x;y)$ as defined above. So we want to compute
\begin{align} \label{int boundary step 3}
\frac{1}{(2\pi)^{m+n}}\int\limits_{C_{1,1}}(x;q)\varphi(q,x)^m \dd\varphi(x,q)^n
\end{align}
for $n\geq 1$ and $m\geq 0$.
\\As before we use Stokes' theorem and find that
\begin{align} \label{result boundary step 3}
\begin{split}
&\int\limits_{C_{1,1}}(x;q)\varphi(q,x)^m \dd\varphi(x,q)^n=\int\limits_{\substack{C_{1,1}\\\mathrm{Re}(x)>q}}\varphi(q,x)^m \dd\varphi(x,q)^n
\\&=\varphi(q,y_-)^m\varphi(y_-,q)^n-\varphi(q,t)^m\varphi(t,q)^n=2^m\pi^{m+n}
\end{split}
\end{align} 
for all $m\geq 0$ and all $n\geq 1$.

\subsubsection{Putting everything together}
Now we can use the results from steps $1$-$3$ to compute the Kontsevich weight (\ref{Kontsevich weight family 2}) of the graphs $\Upsilon_n$, $n\geq 0$, described at the beginning of Section \ref{subsec: one boundary vertex}. Integrating over $z_i$ for $i=1,\ldots,n$, and applying (\ref{result boundary wedges new}), we get
\begin{align} \label{Kontsevich weight family 2 expanded better}
\begin{split}
w_{\Upsilon_n}&=\frac{1}{(2\pi)^{2n+1}}\int\limits_{C_{n+1,1}}\dd\varphi(x,q)\dd\varphi(z_1,x)\dd\varphi(z_1,q)\cdots \dd\varphi(z_n,x)\dd\varphi(z_n,q)
\\&=\frac{1}{(2\pi)^{n+1}}\int\limits_{C_{1,1}}\big(\varphi(x,q)-\varphi(q,x)-\pi+2\pi(x;q)\big)^{n}\dd\varphi(x,q)
\\&=\frac{1}{(2\pi)^{n+1}}\sum\limits_{k=0}^{n}\sum\limits_{l=0}^{n-k}\sum\limits_{s=0}^{n-k-l}{n\choose k}{n-k\choose l}{n-k-l\choose s}(-1)^{l+s}
\\&\times \int\limits_{C_{1,1}}\varphi(x,q)^{n-k-l-s}\varphi(q,x)^{s}\pi^{l}(2\pi(x;q))^{k}\dd\varphi(x,q)
\\&=\sum\limits_{k=0}^{n}\sum\limits_{l=0}^{n-k}\sum\limits_{s=0}^{n-k-l}{n\choose k}{n-k\choose l}{n-k-l\choose s}\frac{(-1)^{l+s}}{2^{n-k+1}\pi^{n-k-l+1}(n-k-l-s+1)}
\\&\times\int\limits_{C_{1,1}}(x;q)^{k}\varphi(q,x)^{s}\dd\varphi(x,q)^{n-k-l-s+1}.
\end{split}
\end{align}
We note that for $k=0$ we have
\begin{align}
\int\limits_{C_{1,1}}\varphi(q,x)^{s}\dd\varphi(x,q)^{n-l-s+1}=
\begin{cases}
(2\pi)^{n-l+1}, & \text{if }s=0 \\ 2^s\pi^{n-l+1}, & \text{if } s>0
\end{cases}
\end{align}
where we have used (\ref{result boundary step 2}). Similarly, for $k\geq 1$ we have
\begin{align}
\int\limits_{C_{1,1}}(x;q)^{k}\varphi(q,x)^{s}\dd\varphi(x,q)^{n-k-l-s+1}=\int\limits_{C_{1,1}}(x;q)\varphi(q,x)^{s}\dd\varphi(x,q)^{n-k-l-s+1}=2^s\pi^{n-k-l+1},
\end{align}
where we have used (\ref{result boundary step 3}).
\\Plugging the above two results into the last line of (\ref{Kontsevich weight family 2 expanded better}) we get
\begin{align} \label{intermediate step family 2}
\begin{split}
w_{\Upsilon_n}&=\underbrace{\sum\limits_{k=0}^{n}\sum\limits_{l=0}^{n-k}\sum\limits_{s=0}^{n-k-l}{n\choose k}{n-k\choose l}{n-k-l\choose s}\frac{(-1)^{l+s}}{2^{n-k-s+1}(n-k-l-s+1)}}_{=:A(n)}
\\&-\underbrace{\sum\limits_{l=0}^{n}{n\choose l}\frac{(-1)^l}{2^{n+1}(n-l+1)}}_{=:B(n)}+\underbrace{\sum\limits_{l=0}^{n}{n\choose l}\frac{(-1)^l}{2^{l}(n-l+1)}}_{=:C(n)}.
\end{split}
\end{align}
As shown in Appendix \ref{Appendix:Binomial sums}, we have that
\begin{align} \label{Binomial Sums results}
\begin{split}
A(n)&=\frac{(-1)^{n}}{2^{n+1}(n+1)},
\\B(n)&=\frac{(-1)^{n}}{2^{n+1}(n+1)},
\\C(n)&=\frac{1+(-1)^{n}}{2^{n+1}(n+1)}.
\end{split}
\end{align}
Hence, we finally have
\begin{align} \label{final result family 2}
w_{\Upsilon_n}=\frac{1+(-1)^{n}}{2^{n+1}(n+1)}, \qquad n\geq 0.
\end{align}
In particular, we see that
\begin{align} 
w_{\Upsilon_n}=0 \qquad\text{for odd }n\geq 1.\;
\end{align}
The Kontsevich weights of the first few graphs are given in Table \ref{table: Kontsevich weights 2} below.
\begin{table}[ht] 
\begin{center}
 \begin{tabular}{|c|cccccccccc|} 
 \hline
 $n$ & $0$ & $1$ & $2$ & $3$ & $4$ & $5$ & $6$ & $7$ & $8$ & $9$ \\ 
 \hline
 \\[-0.35cm]
$w_{\Upsilon_n}$ & $1$ & $0$ & $\frac{1}{12}$ & $0$ & $\frac{1}{80}$ & $0$ & $\frac{1}{448}$ & $0$ & $\frac{1}{2304}$ & $0$  \\[0.08cm] 
\hline 
\end{tabular}
\end{center}
\caption{Kontsevich weights of the graphs $\Upsilon_n$ for $n=0,1,\ldots,9$}
\label{table: Kontsevich weights 2}
\end{table}

As a sanity check we have the following: For $n=0$ the graph $\Upsilon_0$ is just a single edge as in Figure \ref{fig:Kontsevich weight 6}(b) and its weight is zero according to \cite[Section 6.4.3]{K}. For $n=1$ the graph $\Upsilon_1$ is just a single edge with one wedge attached as in Figure \ref{fig:Kontsevich weight 7}(b) and its weight is $0$ according to \cite[Appendix B]{FelderWillwacher2009}. For $n=2$ the graph $\Upsilon_2$ is a single edge with two wedges attached as in Figure \ref{fig:Kontsevich weight 7}(c) and its weight is $\frac{1}{12}$ according to \cite[Appendix A]{VanBergh2009}. So at least for $n=0,1,2$ our formula (\ref{final result family 2}) for the Kontsevich weights $w_{\Gamma_n}$ produces the correct values.

\subsection{Case 3: Two Boundary Vertices} \label{subsec: two boundary vertices}
\begin{figure}[ht] 
 \centering
 \includegraphics[width=0.25\textwidth]{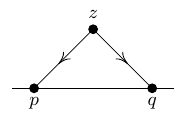}
 \caption{Graphs in the case $(r,m)=(0,2)$ consist of wedges attached to the two boundary vertices}
 \label{fig:Kontsevich weight 8}
\end{figure}

Finally, we will treat the case $(r,m)=(0,2)$, i.e. the case where we have two boundary vertices and no $R$-vertex. In that case we get a family of graphs $(\Lambda_n)_{n\geq 0}$, where $\Lambda_n$ is the graph with $n$ wedges as in Figure \ref{fig:Kontsevich weight 8} (stemming from $n$ $\mathsf{T}\phi^{*}_{x}\pi$-vertices) attached to the two boundary vertices.

\begin{figure}[ht] 
 \centering
 \includegraphics[width=0.85\textwidth]{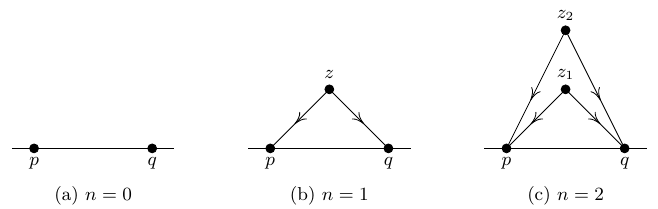}
 \caption{Graphs $\Lambda_n$ for (a) $n=0$, (b) $n=1$ and (c) $n=2$ wedges attached to the two boundary vertices}
 \label{fig:Kontsevich weight 9}
\end{figure}

Examples of the graphs $\Lambda_n$ are given in Figure \ref{fig:Kontsevich weight 9} below for $n=0,1,2$.

The Kontsevich weight of the graph $\Lambda_n$ for $n\geq 0$ is given by 
\begin{align} \label{Kontsevich weight family 3}
w_{\Lambda_n}=\frac{1}{(2\pi)^{2n}}\int\limits_{C_{n,2}}\dd\varphi(z_1,p)\dd\varphi(z_1,q)\cdots \dd\varphi(z_n,p)\dd\varphi(z_n,q).
\end{align}

\begin{rem}
For $n=0$ we simply set $\dd\varphi(z_1,p)\dd\varphi(z_1,q)\cdots \dd\varphi(z_n,p)\dd\varphi(z_n,q)=1$ in the integral above.
\end{rem}

\begin{rem}
As before, the ordering of the edges of the graph $\Lambda_n$ specified in (\ref{Kontsevich weight family 3}) above determines the sign of $w_{\Lambda_n}$. Throughout this whole section we will stick to this ordering. 
\end{rem}

\begin{rem} \label{rmk: conf spaces}
Since we work with the configuration space $\mathrm{Conf}_{n,m}$ as in \eqref{eq:conf}, and in particular with the quotient $C_{n,m}=\mathrm{Conf}_{n,m}/G^{(1)}$, it follows that $C_{0,2}$ is a single point (and not a two-element set).
\end{rem}

Finally, our goal is to compute (\ref{Kontsevich weight family 2}) explicitly for the given family of graphs. However, this time the computation is much easier and shorter than before.
\\For the boundary vertices $p,q,\in\mathbb{R}$ with $p<q$ we have already computed the Kontsevich weight of a wedge as in Figure \ref{fig:Kontsevich weight 8} at the end of Section \ref{subsubsec step 1 family 1}. Our result was
\begin{align} \label{result 2 boundary wedge}
\frac{1}{(2\pi)^2}\int\limits_{C_{1,2}}\dd\varphi(z,p)\dd\varphi(z,q)=\frac{1}{2}.
\end{align}
For the sake of completeness and to make sure that we get the same result, let us nonetheless do a direct computation. For a wedge as in Figure \ref{fig:Kontsevich weight 8}, we compute the corresponding integral
\begin{align}
\frac{1}{(2\pi)^2}\int\limits_{z\in\mathbb{H}}\dd\varphi(z,p)\dd\varphi(z,q),
\end{align}
with $p,q\in\mathbb{R}$, $p<q$ fixed. As before, we use Stokes' theorem
\begin{align}
\int\limits_{z\in\mathbb{H}}\dd\varphi(z,p)\dd\varphi(z,q)=\int\limits_{\partial}\varphi(z,p)\dd\varphi(z,q),
\end{align}
where $\partial$ is the boundary of the integration domain depicted in Figure \ref{fig:Kontsevich integral 5} below.

\begin{figure}[ht] 
 \centering
 \includegraphics[width=1.0\textwidth]{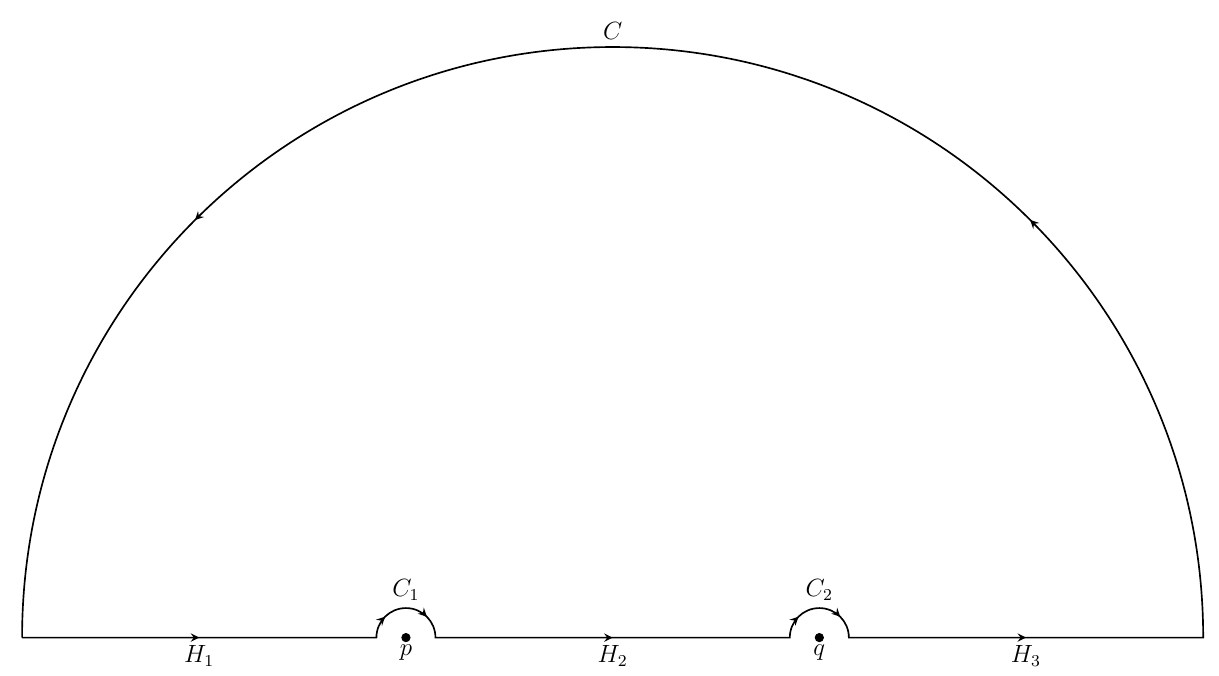}
 \caption{Boundary $\partial$ of the integration domain: $C$ is the half-circle at infinity, the half circles $C_1$ and $C_2$ have infinitesimal radius and $H_{1}\cup H_{2}\cup H_{3}$ is the real line.}
 \label{fig:Kontsevich integral 5}
\end{figure} 

As usual, let us have a look at the different boundary components:
\begin{itemize}
\item On $H_{1}\cup H_{2}\cup H_{3}$: $z\in\mathbb{R}$ and hence $\dd\varphi(z,q)=0$
\item On $C_1$: $z=p+\varepsilon \mathrm{e}^{-i\theta}$ for $\varepsilon\rightarrow 0$ $\implies$ $\dd\varphi(z,q)=\dd\text{arg}(1)=0$
\item On $C_2$: $z=q+\varepsilon \mathrm{e}^{-i\theta}$ for $\varepsilon\rightarrow 0$ $\implies$ $\varphi(z,p)\rightarrow\varphi(q,p)=0$
\item On $C$: $z=R\mathrm{e}^{i\theta}$ for $R\rightarrow\infty$ and $\theta\in[0,\pi]$ $\implies$ $\varphi(z,p)=\varphi(z,q)=2\theta$
\end{itemize}
We can then compute the integral
\begin{align}
\int\limits_{z\in\mathbb{H}}\dd\varphi(z,p)\dd\varphi(z,q)=\int\limits_{\partial}\varphi(z,p)\dd\varphi(z,q)=\int\limits_{0}^{\pi}4\theta d\theta=2\pi^{2},
\end{align}
which indeed agrees with (\ref{result 2 boundary wedge}) after dividing by $(2\pi)^2$.
\\With this result at hand, it is now easy to compute the Kontsevich weight of the graph $\Lambda_n$ for $n\geq 1$. We get
\begin{align} \label{Kontsevich weight family 3 final}
w_{\Lambda_n}=\frac{1}{(2\pi)^{2n}}\int\limits_{C_{n,2}}\dd\varphi(z_1,p)\dd\varphi(z_1,q)\cdots \dd\varphi(z_n,p)\dd\varphi(z_n,q)=\frac{1}{(2\pi)^{2n}}(2\pi^{2})^{n}=\frac{1}{2^n}.
\end{align}
For $n=0$ it is not hard to see that
\begin{align}
w_{\Lambda_0}=\int\limits_{C_{0,2}}1=1,
\end{align}
since $C_{0,2}$ is a single point (cf Remark \ref{rmk: conf spaces}). So all in all we finally have
\begin{align} \label{final result family 3}
w_{\Lambda_n}=\frac{1}{2^n},\qquad n\geq 0.
\end{align}

\subsection{Another Approach for the Explicit Computation of $w_{\Gamma_n}$ and $w_{\Upsilon_n}$}
\label{subsec:another_explicit_approach}
We want to give a more fast and explicit approach for the computation of the weights $w_{\Gamma_n}$ and $w_{\Upsilon_n}$. The following approach has the advantage that it doesn't require the use of the hypergeometric function for $w_{\Gamma_n}$ but rather gives an explicit expression in terms of $w_{\Upsilon_n}$.
First note that 
\begin{equation}
    \dd\varphi(z,x)=\frac{1}{4\pi i}\dd\log\left(\frac{(z-x)(z-\bar x)}{(\bar z-x)(\bar z-\bar x)}\right), \qquad \forall z,x\in \mathbb{H}\cup \R.
\end{equation}
Moreover, recall from \cite[Lemma 5.3]{BanksPanzerPym2018} the formula
\begin{equation}
\label{eq:integral1}
    \int_{z\in \mathbb{H}}\dd\varphi(z,x)\land \dd\varphi(z,y)=\frac{1}{2\pi i}\log\left(\frac{x-\bar y}{y-\bar x}\right),\qquad\forall x,y\in \mathbb{H}\cup \R.
\end{equation}
Integrating out the $z_i$ variables in $w_{\Upsilon_n}$ using \eqref{eq:integral1}, we get 
\begin{equation}
\label{eq:integral2}
    w_{\Upsilon_n}=\int_{C_{1,1}}\left(\frac{1}{2\pi i}\log\left(\frac{x-p}{p-\bar x}\right)\right)^n\frac{1}{2\pi i}\dd\log\left(\frac{x-p}{\bar x-p}\right)=\frac{1}{n+1}\int_{C_{1,1}}\dd\left(\frac{1}{2\pi i}\dd\log\left(\frac{x-p}{p-\bar x}\right)\right)^{n+1}
\end{equation}
because $\dd\log\left(\bar x-p\right)=\dd\log\left(p-\bar x\right)$. Now recall that $C_{1,1}=(\mathbb{H}\times \R)/(\R^{>0}\ltimes \R)$ is isomorphic to $\R$; for instance every point in the quotient can be represented uniquely by a pair $(i,p)$ with $i$ the imaginary unit and $p\in \R$. Hence we get
\begin{align}
\begin{split}
    w_{\Upsilon_n}&=\frac{1}{n+1}\int_{-\infty}^\infty\frac{\dd}{\dd p}\left(\frac{1}{2\pi i}\log\left(\frac{i-p}{p+i}\right)\right)^{n+1}\dd p\\
    &=\left(\frac{1}{2\pi i}\log\left(\frac{i-p}{p+i}\right)\right)^{n+1}\Bigg\vert_{p={-\infty}}^{p=\infty}\\
    &=\frac{1+(-1)^n}{2^{n+1}(n+1)},
\end{split}
\end{align}
where we have used the boundary values
\begin{equation}
    \lim_{p\to \pm\infty}\log\left(\frac{i-p}{p+i}\right)=\lim_{p\to\pm\infty}\log\left(-1+\frac{2 i}{p}+O(1/p^2)\right)=\pm i\pi.
\end{equation}

Now, using \eqref{eq:integral1} and integrating out the $z_i$ variables of $w_{\Gamma_n}$, we get 
\begin{equation}
    w_{\Gamma_n}=\int_{C_{2,0}}\left(\frac{1}{2\pi i}\log\left(\frac{x-\bar y}{z-\bar x}\right)\right)^n\dd\varphi(x,y)\land \dd\varphi(y,x).    
\end{equation}
Note that $\dd\varphi(x,y)=\dd\varphi(y,x)+\frac{1}{2\pi i}\log\left(\frac{x-\bar y}{y-\bar x}\right)$, and thus
\begin{align}
\begin{split}
    w_{\Gamma_n}&=\int_{C_{2,0}}\left(\frac{1}{2\pi i}\log\left(\frac{x-\bar y}{y-\bar x}\right)\right)^n\frac{1}{2\pi i}\dd\log\left(\frac{x-\bar y}{y-\bar x}\right)\land \dd\varphi(y,x)\\
    &=\frac{1}{n+1}\int_{C_{2,0}}\dd\left[\left(\frac{1}{2\pi i}\log\left(\frac{x-\bar y}{y-\bar x}\right)\right)^{n+1} \dd\varphi(y,x)\right].
\end{split}
\end{align}
By Stokes' theorem, the integral is reduced to the boundary of Kontsevich's eye $\bar C_{2,0}$ \cite[Section 5.2.]{K}. This boundary has three components:
\begin{itemize}
    \item The iris of the eye which is isomorphic to $S^1$, corresponding to the collision $x\to y$. This gives zero contribution, since $\log\left(\frac{x-\bar y}{y-\bar x}\right)=0$ for $x=y$. 
    \item The upper eyelid, corresponding to the limit $\vert x\vert\to \infty$. This gives no contribution, since $\dd\varphi(y,x)=0$ in the limit. 
    \item The lower eyelid, corresponding to the collision $x\to \bar x$ onto the real line. 
\end{itemize}
The lower eyelid is isomorphic to $C_{1,1}$ and by comparison with the integrand of $w_{\Upsilon_{n+1}}$ in \eqref{eq:integral2} with $p:=x=\bar x$, we get 
\begin{equation}
    w_{\Gamma_n}=\frac{1}{n+1}w_{\Upsilon_{n+1}}=\frac{1-(-1)^n}{2^{n+2}(n+1)(n+2)}.
\end{equation}
One can easily check that this formula produces the same values as in Table \ref{table: Kontsevich weights 1}.

\section{Including the Weights} 
\label{sec:def. quant of symplectic poisson mnfs}

\subsection{The Product $P(\mathsf{T}\phi^*\pi)$}
Let $\sigma,\tau\in\Gamma(\Hat{\Sym}(T^*M)[[\hbar]])$ be sections and let $x\in M$. Using the Kontsevich weights computed above, we get
\begin{equation} \label{star product explicit}
P(\mathsf{T}\phi^*_x\pi)(\sigma_x \otimes \tau_x )=\sum\limits_{n=0}^{\infty}\frac{\hbar^n}{2^{2n}n!}(\mathsf{T}\phi^*_x\pi)^{i_1 j_1}\cdots(\mathsf{T}\phi^*_x\pi)^{i_n j_n}(\sigma_x)_{\,,i_1\cdots i_n}(\tau_x)_{\,,j_1\cdots j_n}
\end{equation}
where we sum over all the indices $i_1,\ldots,i_n,j_1,\ldots,j_n$. Moreover, we use the notation, where the indices on the right of the comma denote derivatives with respect to the corresponding variable, e.g.
\begin{equation}
    R_{i,i_1\dotsm i_k}:=\de_{i_1}\dotsm \de_{i_k}R_i.
\end{equation}

\subsection{The Connection 1-form $A(R,\mathsf{T}\phi^*\pi)$}
In Section \ref{subsec: one boundary vertex} we have obtained the Kontsevich weights
\begin{align} \label{final result family 2 second}
w_{\Upsilon_n}=\frac{1+(-1)^{n}}{2^{n+1}(n+1)}, \qquad n\geq 0
\end{align}
of the family of graphs $(\Upsilon_n)_{n\geq 0}$. Let $\sigma\in\Gamma(\Hat{\Sym}(T^*M)[[\hbar]])$ be a section and fix $x\in M$. For $R$ as in Section \ref{subsec:formal_geometry} we set $R_x(y):=R(x,y)$ and $(R_x)^k_i(y):=R^k_i(x,y)$. Using the Kontsevich weights above, we get
\begin{align}
\begin{split}
A(R_x,\T\phi^*_x\pi)(\sigma_x)&=\dd x^iA\left((R_x)^k_{i}\frac{\partial}{\partial y^k},\T\phi^*_x\pi\right)(\sigma_x)
\\&=\dd x^i\sum\limits_{n=0}^{\infty}\frac{\hbar^n}{2^n n!}\frac{1+(-1)^{n}}{2^{n+1}(n+1)}(\T\phi^*_x\pi)^{i_1 j_1}\cdots(\T\phi^*_x\pi)^{i_n j_n}(R_x)^k_{i,i_1\cdots i_n}(\sigma_x)_{\,,kj_1\cdots j_n}
\end{split}
\end{align}
where we again sum over all indices $i,k,i_1,\ldots,i_n,j_1,\ldots,j_n$. 
\\This allows us to write down an explicit expression for the deformed Grothendieck connection, namely
\begin{align} \label{deformed Grothendieck explicit}
\begin{split}
(\mathcal{D}_G)_x&=\dd_x+A(R_x,\T\phi^*_x\pi)=\left(\frac{\partial}{\partial x^i}+A\big((R_x)_i,\T\phi^*_x\pi\big)\right)\dd x^i
\\&=\left(\frac{\partial}{\partial x^i}+\sum\limits_{n=0}^{\infty}\frac{\hbar^n}{2^n n!}\frac{1+(-1)^{n}}{2^{n+1}(n+1)}(\T\phi^*_x\pi)^{i_1 j_1}\cdots(\T\phi^*_x\pi)^{i_n j_n}(R_x)^k_{i,i_1\cdots i_n}\frac{\partial^{n+1}}{\partial y^{j_n}\cdots\partial y^{j_1}\partial y^k}\right)\dd x^i
\end{split}
\end{align}

\subsection{The Curvature 2-form $F(R,R,\mathsf{T}\phi^*\pi)$}
In Section \ref{subsec: no boundary vertices} we have obtained the Kontsevich weights in terms of the hypergeometric function and gave a more explicit formula in Section \ref{subsec:another_explicit_approach}
\begin{equation} \label{Kontsevich weight whole expression 2 second}
w_{\Gamma_n}=\frac{1-(-1)^n}{2^{n+2}(n+1)(n+2)},\qquad n\geq 0
\end{equation}
for the family of graphs $(\Gamma_n)_{n\geq 0}$. Using these weights above we then get for $x\in M$ 
\begin{multline} \label{Weyl curvature explicit}
F(R_x,R_x,\T\phi^*_x\pi)=\dd x^i\wedge \dd x^j F\big((R_x)_i,(R_x)_j,\T\phi^*_x\pi\big)
\\=\dd x^i\wedge \dd x^j\sum\limits_{n=0}^{\infty}\frac{\hbar^n}{2^n n!}\frac{1-(-1)^n}{2^{n+2}(n+1)(n+2)}(\T\phi^*_x\pi)^{i_1 j_1}\cdots(\T\phi^*_x\pi)^{i_n j_n}(R_x)^k_{i,l i_1\cdots i_n}(R_x)^l_{j,k j_1\cdots j_n}
\end{multline}
where, as usual, we sum over the indices $i,j,k,l,i_1,\ldots,i_n,j_1,\ldots,j_n$. 

\subsection{A Fedosov-type Equation for Poisson Manifolds} 
We can now write down the \emph{modified deformed Grothendieck connection} as defined in \cite{CF3,CMW3}, namely
\begin{align}
\overline{\mathcal{D}}_G=\mathcal{D}_G+[\gamma,\enspace]_\star,
\end{align}
where the deformed Grothendieck connection $\mathcal{D}_G$ is explicitly given by \eqref{deformed Grothendieck explicit}, the star product is explicitly given by \eqref{star product explicit}, $[\enspace,\enspace]_\star$ denotes the star commutator and $\gamma\in\Omega^1(M,\Hat{\Sym}(T^*M)[[\hbar]])$ is such that the following Fedosov-type equation holds:
\begin{align} \label{equation for gamma}
F^M+\mathcal{D}_G\gamma+\gamma\star\gamma=0,
\end{align}
with Weyl curvature $F^M=F(R,R,\T\phi^*\pi)$ explicitly given by \eqref{Weyl curvature explicit}. This equation appears in the globalization construction for deformation quantization of Poisson manifolds.
The existence of such a $\gamma$ was given in \cite{CF3,CattaneoFelderTomassini2002}. Let us emphasize a bit more on this existence result. 
Since $\gamma$ takes values in $\Hat{\Sym}(T^*M)[[\hbar]]$ we may write
\begin{align}
\gamma=\gamma_0+\hbar\gamma_1+\hbar^2\gamma_2+\ldots
\end{align} 
Similarly, for the deformed Grothendieck connection we may write
\begin{align}
\mathcal{D}_G=D_G+\hbar^2\mathcal{D}_2+\hbar^4\mathcal{D}_4+\ldots
\end{align}
where $D_G=\dd+L_R$ is the classical Grothendieck connection and where we have used that the Kontsevich weights \eqref{final result family 2 second} satisfy $w_{\Upsilon_0}=1$ and $w_{\Upsilon_n}=0$ for all odd $n\geq 1$.
\\Finally, for the curvature we can write
\begin{align}
F^M=\hbar F_1+\hbar^3 F_3+\hbar^5 F_5+\ldots
\end{align}
where we have used that that the Kontsevich weights \eqref{Kontsevich weight whole expression 2 second} satisfy $w_{\Gamma_n}=0$ for all even $n\geq 0$.
\\This now allows us to decompose Equation \eqref{equation for gamma} into a system of equations depending on the order of $\hbar$. 
\\In order $\hbar^0$ we get the equation
\begin{align} \label{eq:gamma order 0}
D_G\gamma_0=0,
\end{align}
which, according to Section \ref{subsec:formal_geometry}, can be solved by $\gamma_0=\T\phi^* f$ for some smooth function $f\in C^{\infty}(M)$, since the cohomology of the classical Grothendieck connection $H^\bullet_{D_G}(\Gamma(\Hat{\Sym}(T^*M)))$ is concentrated in degree zero (by the Poincar\'e Lemma) and 
\begin{equation}
H^0_{D_G}(\Gamma(\Hat{\Sym}(T^*M)))\cong \T\phi^*C^\infty(M)\cong C^\infty(M). 
\end{equation}
\\In order $\hbar^1$ we get the equation
\begin{align} \label{eq:gamma order 1}
F_1+D_G\gamma_1+(\gamma_0\star\gamma_0)_1=0.
\end{align}
Using the Bianchi identity we see that $D_G F_1=0$ and by Equation \eqref{eq:gamma order 0} it also immidiately follows that $D_G(\gamma_0\star\gamma_0)_1=0$. So we get that $D_G\gamma_1$ is equal to a $D_G$-closed form, but the corresponding cohomology group is trivial. Hence it follows that $D_G\gamma_1$ is equal to a $D_G$-exact form, and thus it is possible to find a $\gamma_1$ that solves Equation \eqref{eq:gamma order 1}. 
\\By induction, one can show that in each order $\hbar^k$ for $k\geq 1$, $D_G\gamma_k$ is equal to a $D_G$-closed and hence $D_G$-exact form depending on the lower order coefficients of $F^M$ and $\gamma$. In particular it follows that there exists a $\gamma_k$ solving the equation for the corresponding order. 

\begin{rem}
Note that in order to globalize Kontsevich's star product one may be tempted to define a bullet product
\begin{equation} \label{bullet product}
(f\bullet g)(x):=\big(P(\T\phi^*\pi)(\T\phi^* f\otimes \T\phi^* g)\big)(x;0).
\end{equation}
This is indeed a well-defined global product on $C^{\infty}(M)[[\hbar]]$, but it is in general not associative. To make this product associative one has to introduce a quantization map (see e.g. \cite{CF3})
\begin{equation}
    \rho\colon H^0_{D_G}(\Gamma(\Hat{\Sym}(T^*M)))\to H^0_{\overline{\calD}_G}(\Gamma(\Hat{\Sym}(T^*M)[[\hbar]]))
\end{equation}
which then again leads to the global star product 
\begin{equation} \label{def:star product global}
f\star_{_M} g:=\left(\rho^{-1}\big(\rho(\T\phi^* f)\star\rho(\T\phi^* g)\big)\right)\Big\vert_{y=0}.
\end{equation}
Here $\star$ denotes Kontsevich's star product and $\star_M$ its global version on $M$.
Using the weights, we can also get an explicit expression for the bullet product \eqref{bullet product} by
\begin{multline}
\big(P(\mathsf{T}\phi^*\pi)(\mathsf{T}\phi^* f\otimes \mathsf{T}\phi^* g)\big)(x;0)
\\=\left(\sum\limits_{n=0}^{\infty}\frac{\hbar^n}{2^{2n}n!}(\mathsf{T}\phi^*_x\pi)^{i_1 j_1}\cdots(\mathsf{T}\phi^*_x\pi)^{i_n j_n}(\mathsf{T}\phi^*_x f)_{\,,i_1\cdots i_n}(\mathsf{T}\phi^*_x g)_{\,,j_1\cdots j_n}\right)(0).
\end{multline}
\end{rem}

\subsection{The Lifted Curvature 2-form $F(\overline{R},\overline{R},\mathsf{T}\overline{\phi}^*\pi)$}
Let $M$ be a smooth manifold and let $\phi\colon TM\rightarrow M$ be a formal exponential map and consider the lift $\overline{\phi}\colon TN\rightarrow N$ to the cotangent bundle $N=T^{*}M$. We set $x=(q,p)\in N$ and $y=(\bar{q},\bar{p})\in T_x N$. Note that this is a particular case of a canonical symplectic manifold.
\\We will consider the lifted vector fields $\overline{R}$ to the cotangent case, which induce lifted interaction vertices within the Feynman graphs which appear in the computation of the connection $1$-form and its curvature $2$-form and see how these terms simplify. First we note that $A(\overline{R},\T\overline{\phi}^*\pi)$ is still given by 
\begin{align}
A(\overline{R}_x,\T\overline{\phi}^*_x\pi)(\sigma_x)=\dd x^i\sum\limits_{n=0}^{\infty}\frac{\hbar^n}{2^n n!}\frac{1+(-1)^{n}}{2^{n+1}(n+1)}(\T\overline{\phi}^*_x\pi)^{i_1 j_1}\cdots(\T\overline{\phi}^*_x\pi)^{i_n j_n}(\overline{R}_x)^k_{i,i_1\cdots i_n}(\sigma_x)_{\,,kj_1\cdots j_n}.
\end{align}
The simplification in this case is a small one: All summands containing a term $(\overline{R}_x)^k_{i,i_1\cdots i_n}$ with more than one derivative with respect to $\bar{p}$ will vanish \cite{Mosh1}. 
\\For the case of the curvature $2$-form $F^N$ the simplification is more interesting. Since for each non-vanishing coefficient $(\T\overline{\phi}^*_x\pi)^{ij}$ one of the two outgoing edges is always representing a $\bar{q}$-derivative and the other corresponding edge representing a $\bar{p}$-derivative (since we work with Darboux coordinates around $x\in N$), we see that the sum in \eqref{Weyl curvature explicit} terminates at $n=2$. Or put differently, we only have to consider the graphs $\Gamma_n$ up to $n=2$, i.e. with at most two wedges attached to the wheel consisting of two $\overline{R}$-vertices (cf. Figure \ref{fig:Kontsevich weight 1} in Section \ref{subsec: no boundary vertices}). Moreover, since the Kontsevich weights \eqref{Kontsevich weight whole expression 2 second} are, up to $n=2$, given by $w_{\Gamma_0}=0$, $w_{\Gamma_1}=\frac{1}{24}$ and $w_{\Gamma_2}=0$, we get
\begin{align} \label{Weyl curvature simplified}
F^N_x=F(\overline{R}_x,\overline{R}_x,\T\overline{\phi}_x^*\pi)=\frac{\hbar}{48}(\T\overline{\phi}^*_x\pi)^{r s}(\overline{R}_x)^k_{i,l r}(\overline{R}_x)^l_{j,k s}\dd x^i\wedge \dd x^j,
\end{align}
where we sum over the indices $i,j,r,s,k,l$ and where again summands containing a term $(\overline{R}_x)^k_{i,lr}$ with more than one derivative with respect to $\bar{p}$ vanish. So in the case of a cotangent bundle we get a much simpler expression for the Weyl curvature $F^N$.

\begin{appendix}

\section{Binomial Sums} \label{Appendix:Binomial sums}
Here we will treat the binomial sums appearing in the expression (\ref{intermediate step family 2}) and show that we indeed get the results stated in (\ref{Binomial Sums results}). 

\subsection{$B(n)$}
Let us start with
\begin{align}
B(n)=\sum\limits_{l=0}^{n}{n\choose l} \frac{(-1)^l}{2^{n+1}(n-l+1)}.
\end{align}
Using the well known identity
\begin{align} \label{Binomial identity 1}
\sum\limits_{l=0}^{n}(-1)^l{n+1\choose l}=\sum\limits_{l=0}^{n}(-1)^l{n\choose l}\frac{n+1}{n-l+1}=(-1)^n
\end{align}
it immidiately follows that
\begin{align} \label{B(n) final}
B(n)=\frac{(-1)^n}{2^{n+1}(n+1)}.
\end{align}

\subsection{$C(n)$}
Let us continue with 
\begin{align}
C(n)=\sum\limits_{l=0}^{n}{n\choose l} \frac{(-1)^l}{2^{l}(n-l+1)}.
\end{align}
Using the identity (\ref{Binomial identity 1}), we can write
\begin{align}
(n+1)C(n)=\sum\limits_{l=0}^{n}{n+1\choose l} \left(-\frac{1}{2}\right)^l.
\end{align}
Using the Binomial theorem we then find 
\begin{align}
\sum\limits_{l=0}^{n}{n+1\choose l} \left(-\frac{1}{2}\right)^l=\left(\frac{1}{2}\right)^{n+1}-\left(-\frac{1}{2}\right)^{n+1},
\end{align}
and hence
\begin{align} \label{C(n) final}
C(n)=\frac{1+(-1)^n}{2^{n+1}(n+1)}.
\end{align}

\subsection{$A(n)$}
Finally, let us treat the case 
\begin{align}
A(n)=\sum\limits_{k=0}^{n}\sum\limits_{l=0}^{n-k}\sum\limits_{s=0}^{n-k-l}{n\choose k}{n-k\choose l}{n-k-l\choose s}\frac{(-1)^{l+s}}{2^{n-k-s+1}(n-k-l-s+1)}.
\end{align}
Write
\begin{align} \label{A(n) step 1}
A(n)=\sum\limits_{k=0}^{n}\sum\limits_{l=0}^{n-k}{n\choose k}{n-k\choose l}\frac{(-1)^l}{2^{n-k+1}}\sum\limits_{s=0}^{n-k-l}{n-k-l\choose s}\frac{(-2)^{s}}{(n-k-l-s+1)}.
\end{align}
We first treat the innermost sum: Set $m=n-k-l$. Then
\begin{align}
\sum\limits_{s=0}^{n-k-l}{n-k-l\choose s}\frac{(-2)^{s}}{(n-k-l-s+1)}=\sum\limits_{s=0}^{m}{m\choose s}\frac{(-2)^{s}}{(m-s+1)}.
\end{align}
Using ${m+1\choose s}=\frac{m+1}{m-s+1}{m\choose s}$, we find that
\begin{align}
\sum\limits_{s=0}^{m}{m\choose s}\frac{(-2)^{s}}{(m-s+1)}=\frac{1}{m+1}\sum\limits_{s=0}^{m}{m+1\choose s}(-2)^{s}.
\end{align}
Applying the Binomial theorem we get that
\begin{align}
\sum\limits_{s=0}^{m}{m+1\choose s}(-2)^{s}=(-1)^{m+1}\big(1-2^{m+1}\big).
\end{align}
Plugging all of this into (\ref{A(n) step 1}), we find
\begin{align} \label{A(n) step 2}
\begin{split}
A(n)&=\sum\limits_{k=0}^{n}\sum\limits_{l=0}^{n-k}{n\choose k}{n-k\choose l}\frac{(-1)^l}{2^{n-k+1}}\frac{(-1)^{n-k-l+1}}{n-k-l+1}\big(1-2^{n-k-l+1}\big)
\\&=\sum\limits_{k=0}^{n}{n\choose k}\frac{(-1)^{n-k+1}}{2^{n-k+1}}\sum\limits_{l=0}^{n-k}{n-k\choose l}\frac{1}{n-k-l+1}\big(1-2^{n-k-l+1}\big).
\end{split}
\end{align}
Moreover, we have
\begin{align}
\sum\limits_{l=0}^{n-k}{n-k\choose l}\frac{1}{n-k-l+1}=\frac{1}{n-k+1}\sum\limits_{l=0}^{n-k}{n-k+1\choose l}=\frac{1}{n-k+1}\big(2^{n-k+1}-1\big),
\end{align}
and
\begin{align}
\sum\limits_{l=0}^{n-k}{n-k\choose l}\frac{2^{n-k-l+1}}{n-k-l+1}=\frac{1}{n-k+1}\sum\limits_{l=0}^{n-k}{n-k+1\choose l}2^{n-k-l+1}=\frac{1}{n-k+1}\big(3^{n-k+1}-1\big).
\end{align}
Hence
\begin{align} \label{A(n) step 3}
\begin{split}
A(n)&=\sum\limits_{k=0}^{n}{n\choose k}\frac{(-1)^{n-k+1}}{2^{n-k+1}(n-k+1)}\big(2^{n-k+1}-3^{n-k+1}\big)
\\&=(-1)^{n+1}\sum\limits_{k=0}^{n}{n\choose k}\frac{(-1)^{k}}{n-k+1}\left(1-\left(\frac{3}{2}\right)^{n-k+1}\right).
\end{split}
\end{align}
Now
\begin{align}
\sum\limits_{k=0}^{n}{n\choose k}\frac{(-1)^{k}}{n-k+1}=\frac{1}{n+1}\sum\limits_{k=0}^{n}{n+1\choose k}(-1)^{k}=\frac{(-1)^n}{n+1},
\end{align}
and
\begin{align}
\begin{split}
\sum\limits_{k=0}^{n}{n\choose k}\frac{(-1)^{k}}{n-k+1}\Big(\frac{3}{2}\Big)^{n-k+1}&=\frac{(-1)^{n+1}}{n+1}\sum\limits_{k=0}^{n}{n+1\choose k}\left(-\frac{3}{2}\right)^{n-k+1}=\frac{(-1)^n}{n+1}
\\&=\frac{1}{2^{n+1}(n+1)}+\frac{(-1)^n}{n+1}.
\end{split}
\end{align}
Finally, we get 
\begin{align} \label{A(n) final}
A(n)=(-1)^{n+1}\left(\frac{(-1)^{n}}{n+1}-\frac{1}{2^{n+1}(n+1)}-\frac{(-1)^{n}}{n+1}\right)=\frac{(-1)^n}{2^{n+1}(n+1)}.
\end{align}

\end{appendix}



\printbibliography

\end{document}